\documentclass[twocolumn]{aastex63}
\usepackage{natbib, color, amsmath}
\graphicspath{{./figures}}
\newcommand{\alf}{Alfv\'en}
\newcommand{\alfic}{Alfv\'enic}
\newcommand{\alficy}{Alfv\'enicity}
\newcommand\kms{km s$^{-1}$}

\begin{document}

\title{Differentiating the acceleration mechanisms in the slow and {\alfic} slow solar wind}
\newcommand{\CfA}{\affiliation{Center for Astrophysics $\vert$ Harvard \& Smithsonian, 60 Garden Street, Cambridge, MA 02138, USA}}

\newcommand{\UMich}{\affiliation{University of Michigan
Department of Climate \& Space Sciences \& Engineering
2455 Hayward St.
Ann Arbor, MI 48109-2143, USA}}

\newcommand{\GSFCHelio}{\affiliation{Heliophysics Laboratory
NASA Goddard Space Flight Center
8800 Greenbelt Rd.
Greenbelt, MD 20771}}

\newcommand{\MSSL}{\affiliation{Mullard Space Science Laboratory, University College London, Holmbury St. Mary, Dorking, Surrey, RH5 6NT, UK}}
\newcommand{\INAF}{\affiliation{INAF - Institute for Space Astrophysics and Planetology, Rome, Italy }}

\author[0000-0002-8748-2123]{Yeimy J. Rivera}
\CfA

\author[0000-0002-6145-436X]{Samuel T. Badman}
\CfA

\author[0000-0003-1138-652X]{J. L. Verniero} \GSFCHelio

\author[0000-0003-0256-9295]{Tania Varesano}
\affiliation{Southwest Research Institute, Boulder, CO 80302, USA}
\affiliation{Department of Aerospace Engineering Sciences, University of Colorado Boulder, Boulder, CO, USA}

\author[0000-0002-7728-0085]{Michael L. Stevens}
\CfA

\author[0000-0002-5702-5802]{Julia E. Stawarz}
\affiliation{Department of Mathematics, Physics, and Electrical Engineering, Northumbria University, Newcastle upon Tyne, UK}

\author[0000-0002-6903-6832]{Katharine K. Reeves}
\CfA

\author[0000-0001-5956-9523]{Jim M. Raines}
\UMich

\author[0000-0002-7868-1622]{John C. Raymond}
\CfA

\author[0000-0002-5982-4667]{Christopher J. Owen}
\MSSL

\author[0000-0002-4149-7311]{Stefano A. Livi}
\affiliation{Southwest Research Institute, San Antonio, TX 78228, USA}
\UMich

\author[0000-0003-1611-227X]{Susan T. Lepri}
\UMich

\author[0000-0002-9325-9884]{Enrico Landi}
\UMich

\author[0000-0001-5258-6128]{Jasper.~S. Halekas}
\affil{Department of Physics and Astronomy, 
University of Iowa, 
Iowa City, IA 52242, USA}

\author[0000-0002-8475-8606]{Tamar Ervin}
\affiliation{Department of Physics, University of California, Berkeley, Berkeley, CA 94720-7300, USA}
\affiliation{Space Sciences Laboratory, University of California, Berkeley, CA 94720-7450, USA}

\author[0000-0003-4437-0698]{Ryan M. Dewey}
\UMich

\author[0000-0002-7426-7379]{Rossana De Marco}
\INAF

\author[0000-0003-2647-117X]{Raffaella D'Amicis}
\INAF

\author[0000-0002-1628-0276]{Jean-Baptiste Dakeyo}
\affiliation{LESIA, Observatoire de Paris, Universit\'e PSL, CNRS, Sorbonne Universit\'e, Universit\'e de Paris, 5 place Jules Janssen, 92195 Meudon, France}
\affiliation{IRAP, Observatoire Midi-Pyrénées, Universit\'e Toulouse III - Paul Sabatier, CNRS, 9 Avenue du Colonel Roche, 31400 Toulouse, France}

\author[0000-0002-1989-3596]{Stuart D. Bale}
\affil{Physics Department, University of California, Berkeley, CA 94720-7300, USA}
\affil{Space Sciences Laboratory, University of California, Berkeley, CA 94720-7450, USA}

\author[0000-0001-6673-3432]{B. L. Alterman}
\GSFCHelio

\begin{abstract}
In the corona, plasma is accelerated to hundreds of kilometers per second, and heated to temperatures hundreds of times hotter than the Sun's surface, before it escapes to form the solar wind. Decades of space-based experiments have shown that the energization process does not stop after it escapes. Instead, the solar wind continues to accelerate and it cools far more slowly than a freely-expanding adiabatic gas. Recent work suggests that fast solar wind requires additional momentum beyond what can be provided by the observed thermal pressure gradients alone whereas it is sufficient for the slowest wind. The additional acceleration for fast wind can be provided through an {\alf} wave pressure gradient. Beyond this fast-slow categorization, however, a subset of slow solar wind exhibits high {\alf}icity that suggest {\alf} waves could play a larger role in its acceleration compared to conventional slow wind outflows. Through a well-timed conjunction between Solar~Orbiter and Parker Solar Probe, we trace the energetics of slow wind to compare with a neighboring {\alf}ic slow solar wind stream. An analysis that integrates remote and heliospheric properties and modeling of the two distinct solar wind streams finds {\alf}ic slow solar wind behaves like fast wind, where a wave pressure gradient is required to reconcile its full acceleration, while non-{\alf}ic slow wind can be driven by its non-adiabatic electron and proton thermal pressure gradients. Derived coronal conditions of the source region indicate good model compatibility but extended coronal observations are required to effectively trace solar wind energetics below Parker's orbit.

\end{abstract}


\section{Introduction} \label{intro}

With the launch of Parker Solar Probe (Parker; \citealt{Fox2016}) in 2018, the heliospheric community has been able to observe solar wind at distances closer than ever before, extending our view nearer to where the solar wind is actively heated and accelerated. One of the most notable discoveries from Parker was the near-ubiquitous appearance of high-amplitude magnetic field reversals, or magnetic switchbacks, near perihelion passes \citep{Bale2019, Kasper2019}. The switchbacks are {\alfic} in nature \citep{Belcher1971}, characterized by a rapid change in the magnetic field direction, and correlated velocity and magnetic field fluctuations with near constant magnetic field magnitude. Switchbacks are typically observed to cluster together in coherent patches consisting of many individual velocity and magnetic field spikes with sharp discontinuities at their boundaries \citep{Horbury2020}. The spatial scale of patches are of the scale of supergranulation at the Sun and contain compositional signatures suggesting they are linked to the boundaries of coronal holes \citep{Bale2021, Fargette2021, Rivera2024b}. From a statistical analysis coupling Parker, Helios, and Ulysses data, it is found that switchback occurrence with radial distance depends on their duration and size. Smaller duration switchbacks decrease in cumulative counts per kilometer while larger scale switchbacks increase, suggesting a dynamic radial evolution \citep{Tenerani2021, Jagarlamudi2023}. Given the gradual restructuring of switchbacks with increasing heliocentric distance, their decay is attractive process to explain the additional heating and acceleration experienced by the solar wind at the inner heliosphere.

Until recently, a comprehensive assessment of the energy sources and physical mechanisms driving outflow dynamics has only been quantified on a statistical basis. For example, \cite{Halekas2023} examines the energy budget of the solar wind to quantify the role of the proton pressure, ambipolar electric potential, and wave energy in the acceleration of different speed streams. Through statistical observations from Parker across $13.3-100R_{\Sun}$, \cite{Halekas2023} finds that the contribution of wave energy flux becomes increasingly important to explain the acceleration of the fast speed wind. Conversely, the acceleration of the slower speed wind can be captured entirely through the electron thermal gradient (or equivalently, the ambipolar potential) \citep{Halekas2022}. This result strongly suggested that the energy content from {\alf} waves is a critical energy component required to explain the extended heating and acceleration of coronal hole wind. Recent work by \cite{Rivera2024} confirmed that large amplitude {\alf} waves assembled in switchback patches contained the necessary energy to explain the heating and acceleration experienced by a moderately fast wind stream across the inner heliosphere. However, such an analysis has not been carried out for the energetics of the slow speed wind beyond the corona. \cite{Telloni2023} finds the energy partitioning of a slow wind stream (with a predicted 450 km s$^{-1}$ asymptotic speed) in the corona, below $<13R_{\sun}$, indicating that the acceleration is driven by the wave energy flux and electric potential. Our analysis will track the flow of slow wind beyond the corona to quantify the energetics driving its evolution just beyond the {\alf} surface, a location that marks where the solar wind increases above the local {\alf} speed. 

\begin{figure*}[]
	\centering
	\includegraphics[angle=-90, width=0.45\linewidth]{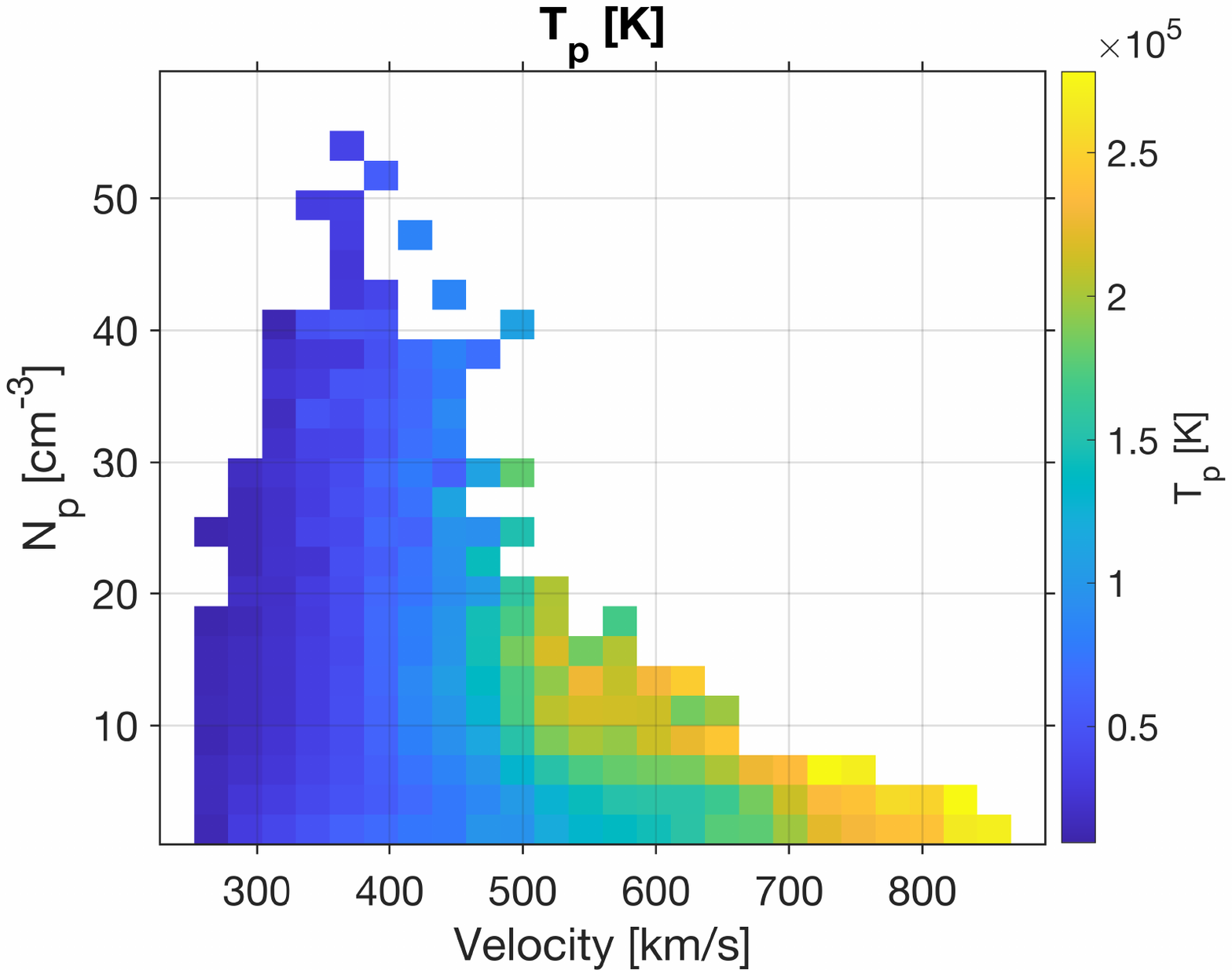}
        \includegraphics[angle=-90, width=0.45\linewidth]{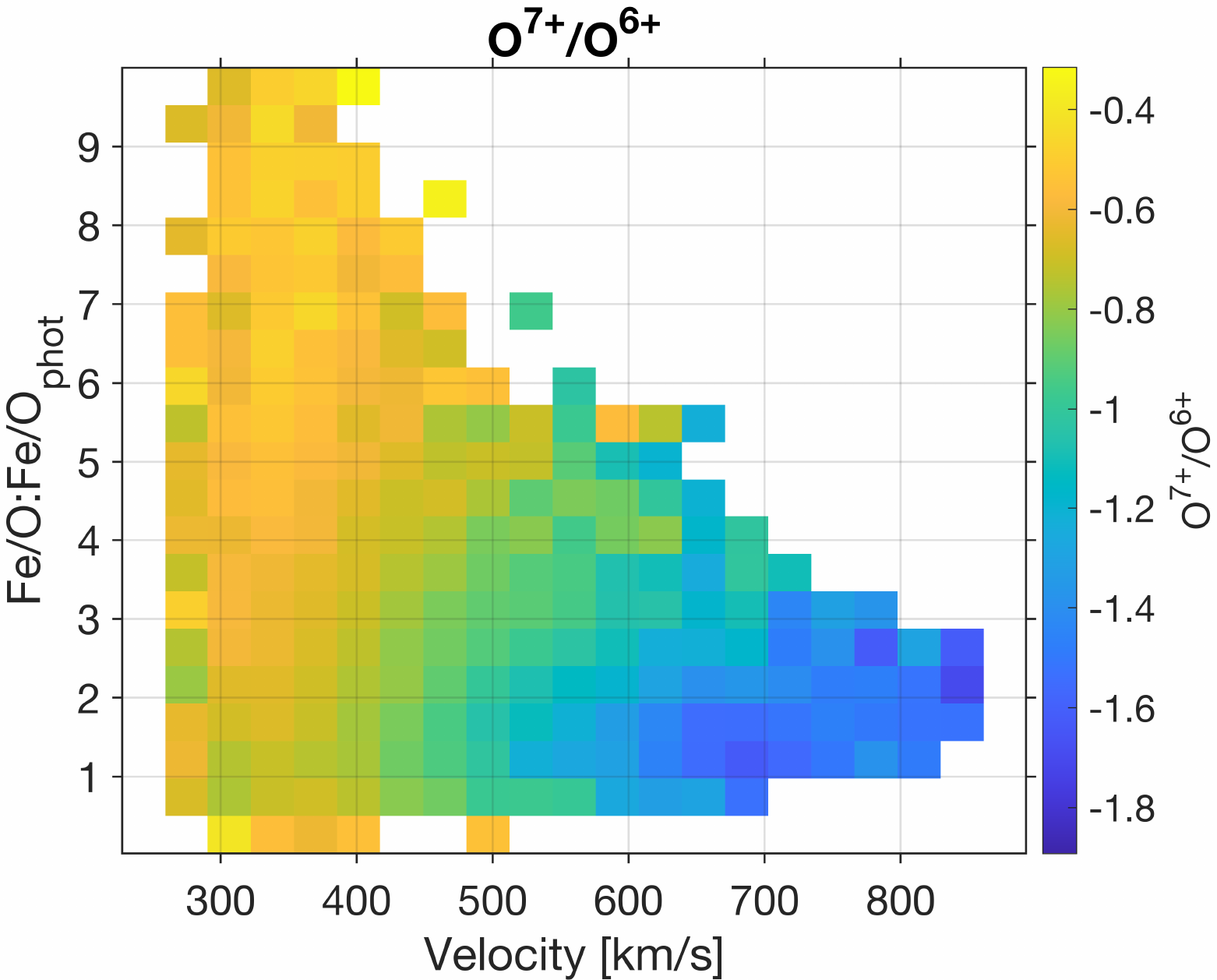}
	\caption{2-D histograms showing the distribution (left) proton density across wind speed where the color indicates the average proton temperature, and (right) Fe/O:Fe/O$_{phot}$, ratio of elemental Fe to O normalized to its photospheric abundances across wind speed where the color indicates the O$^{7+}$/O$^{6+}$ ion ratio.}.
	\label{fig:wind_properties}
\end{figure*}

In particular, the slow speed wind is more variable compared to the fast wind \citep{Abbo2016}, often encompassing properties usually observed in the faster speed wind, e.g. elemental composition \citep{Stakhiv2015, Lepri2013}, coronal temperature as indicated by charge states \citep{Xu_Borovsky2015, Wang2016}, variable mass flux, variation in its {\alficy} as indicated by its cross-helicity \citep{DAmicis2015, Perrone2020, Perrone2022}, helium abundance \citep{Ogilvie1974, Aellig2001, Kasper2007, Kasper2012, Alterman2019, Alterman2021}, which indicate diverse coronal sources \citep{Lynch2023, Baker2023, Ervin2024, Ervin2024b}. The variability of slow solar wind properties is characterized by 2-D histograms in Figure \ref{fig:wind_properties} of the indicated properties from observations taken by the Advanced Composition Explorer
\citep[ACE;][]{Stone1998} collected between 1998--2011. The figure shows the relationship between the different parameters and the variability of the slow wind as compared to the fast speed wind across solar cycle 23 within all parameters. Of particular interest for the present study, recent work has sub-categorized the slow speed wind as {\alf}ic or non-{\alf}ic through its cross-helicity character suggesting the wave energy can vary across solar wind of similar speed \citep{DAmicis2015, DAmicis2019}. 

To effectively investigate the role of the thermal and wave energy fluxes, and their associated thermal pressure and {\alf} wave pressure gradient in connection to the {\alf}ic and non-{\alf}ic slow solar wind dynamics, it is necessary to examine their radial evolution within the stream. Such a connection can only be accomplished during conjunctions between the different spacecraft that align to intersect the same solar wind stream at different stages of its evolution. These kinds of coordinated observations enable the direct assessment of energy transfer over this region, connecting the dynamics and heating to the dissipation of the energy content of the {\alf} waves necessary for the present study \citep{Velli2020}.

To capture the slow solar wind evolution, this work examines a spacecraft co-alignment that sampled the same solar wind streams (as supported by several pieces of evidence including stream mapping and compositional diagnostics) at the sub-{\alf}ic periphery to $\sim0.6$ au, where the escaping solar wind is connected to fully formed solar wind. By connecting the solutions from a Parker solar wind model \citep{Parker1958, Parker1960, Dakeyo2022, Shi2022} constrained by measurements at both spacecraft, as well as at the corona through remote observations, large scale energetics are traced across an expanding flux tube in each stream. We show that the {\alf} waves in the form of switchbacks carry and, ultimately, dissipate energy at the order of magnitude needed to sustain both the temperature gradient and acceleration observed in the {\alf}ic slow solar wind stream. This result strongly suggests that, as observed in the fast solar wind \citep{Rivera2024}, the dissipation of {\alf} wave energy carried by magnetic switchbacks and momentum transferred to the particles continues to power the {\alf}ic slow speed wind across the inner heliosphere. For the non-{\alf}ic slow solar wind, we find it is completely accelerated by the proton and electron thermal pressure gradient in the stream with insignificant contribution from the large-scale wave pressure gradient. Derived temperature and density from remote observations of the {\alfic} slow wind source region show fairly good agreement with modeling results. However, a more rigorous examination of the extended corona, through thoughtful and dedicated multi-spacecraft coordination, is necessary to trace energetics from the low and beyond. We note that details of the {\alf} wave dissipation processes themselves are not examined in this study but are critical to understanding how energy is transferred from larger {\alfic} fluctuations to smaller scales and ultimately dissipated to heat the plasma and partitioned between ions and electrons, e.g. \cite{Gonzalez2021, Adhikari2021, Sioulas2022, Bandyopadhyay2023, Bourouaine2024}.

The paper is organized as follows: Section \ref{sec:streammatching} discusses the observations, conjunction period, ballistic mapping, and solar wind properties. Section \ref{sec:conservationequations} presents the mass, magnetic and energy conservation in the streams. Section \ref{sec:coronal_properties} presents the source region analysis. Section \ref{sec:polytropic} presents the model implementation and results. Section \ref{sec:discussion} and \ref{sec:conclusions} present the discussion and conclusions.

\section{Observations} \label{sec:streammatching}

\subsection{Parker Solar Probe and Solar Orbiter observations}
To compute the stream properties and energetics of the slow and {\alf} slow wind, we examine properties of the solar wind in situ during a conjunction between Parker and Solar Orbiter that occurred in February/March 2022 during Parker's 11th perihelion pass. 

At Parker, we examine observations from the SPAN-Ai instrument to compute proton and alpha number densities, temperatures, and bulk velocities \citep{Livi2022}. SPAN-Ai couples an electrostatic analyzer and time-of-flight (TOF) component to resolve incident angle, mass-per-charge, and energy-per-charge of incoming ions. The mass discrimination made possible by the TOF analyzer allows for the identification of separate ion species in the solar wind, mainly the most abundant species -- protons and alpha particles. SPAN-Ai provides individual 3D particle distributions of proton and alpha particles, at 1.75 second cadence. Due to obstruction of the instrument by the heat shield and solar panels, portions of velocity space are at times unobservable with severity varying according to the direction of flow in the spacecraft frame. We discuss the treatment of the SPAN-Ai observations in the following section. We include electron measurements from SPAN-Ae \citep{Whittlesey2020}. Additionally, the 3D magnetic field components are measured by the fluxgate magnetometer (MAG) on FIELDS at 4 vectors/cycle to capture the rapid changes in the magnetic field \citep{Bale2016}.

At Solar Orbiter, measurements of the magnetic field are taken by another fluxgate magnetometer (MAG) at 8 vectors/second \citep{Horbury2020}. Observations of protons, alpha particles, and heavier ions across this period were taken by the Proton-Alpha System (PAS) and Heavy Ion Sensor (HIS) instruments that are part of the Solar Wind Analyzer (SWA) suite \citep{Owen2020}. The proton and alpha densities, temperatures, and velocities were measured with the PAS instrument with 4s full scan 3D particle distributions. Properties of the alpha population were determined from the PAS measurements using a machine learning statistical clustering technique detailed in \cite{Demarco2023}. Observations of the heavier ions ($Z>2$) were obtained at a 10 minute resolution using the TOF mass spectrometer, HIS \citep{Livi2023}.

\subsection{Coinciding Remote Sensing Observations}
To place a constraint to the modeling results in the low corona, we analyze the source region conditions of the solar wind footpoints. We use the Spectral Imaging of the Coronal Environment (SPICE; \citealt{SPICE2020}) instrument aboard Solar Orbiter and Hinode's EUV Imaging Spectrometer (EIS; \citealt{Culhane2007}) to derive elemental composition, electron density and temperature of the coronal hole identified as the solar wind source region for the intervals of interest in this work. SPICE is a high-resolution spectrometer measuring two EUV wavelength bands between 704--790\AA~ and 973--1049\AA. SPICE observations were taken on 2022-02-23 11:23-10:06UT with a 60s exposure and $2''$ slit covering a 768$\times$630.25$''$ region of the Sun. Near contemporaneous Hinode/EIS observations were taken on 2022-02-23 00:30-01:31UT in two wavelength ranges from 171--211\AA~ to 245--291 \AA~ with a $2''$ slit across a 260$\times$512$''$ FOV with a 60s exposure. The Solar Orbiter position was 6 degrees from the Sun-Earth line where EIS observations were taken. For context, we use the Solar Dynamics Observatory/Atmospheric Imaging Assembly (SDO/AIA; \citealt{Lemen2012}) 193\AA~ image as well as a Full Sun Image (FSI; \citealt{Rochus2020}) image in the 174\AA~ channel from Solar Orbiter. The two rasters overlap in a small region. For the analysis, footpoint mapping of the slow {\alf}ic wind covered a portion of the SPICE FOV and an overlapping region of the EIS FOV as discussed in Section \ref{sec:coronal_properties}.

We note that starting 28 December 2021 to 1 March 2022, there was a problem with the Hinode attitude control system related to maintaining a stable roll angle\footnote{https://solarb.mssl.ucl.ac.uk/JSPWiki/Wiki.jsp?page=EISHistory}. Therefore, the EIS observations in the present work were taken while the instrument was rotated causing the FOV to be 93 degrees counterclockwise rotated as well. The pointing information for the raster was adjusted manually where the Fe {\footnotesize XII} 195.119\AA~ was lined up with features of the AIA 193\AA~ channel (where the Fe {\footnotesize XII} line is dominant), as is typically done. For our purposes, adjustment will not affect the coronal analysis since we are interested in the large scale conditions of the source region. 

\subsection{Stream Matching}
In Parker Encounter 11, the spacecraft was in close Parker spiral alignment with Solar Orbiter for a period centered on 25 February \citep{Rivera2024, Ervin2024, Rivera2024b}. This conjunction is illustrated for the solar co-rotating (Carrington) frame in Figure \ref{fig:relative_pos}. The top and bottom panels show respectively the streams projected into the heliographic equatorial plane, and the trajectories ballistically mapped \citep{Stansby2019, Badman2020} to an altitude of 2.5$R_\odot$. The ballistic mapping utilizes the heliographic location of the relevant spacecraft, and the measured solar wind velocity, to define a Parker spiral field line which approximately traces the path of the relevant plasma parcel back to its origin in the outer corona \citep{Nolte1973, Macneil2022, Koukras2022, dakeyo2024}.

In Figure~\ref{fig:relative_pos}, date labels show how Parker and Solar Orbiter move in opposite heliographic directions, and cross the same streams in reverse order at different heliocentric distances. The x-axis of the right hand panel is the ballistically mapped \emph{Source Surface Longitude} which is used to identify the timeframes of solar wind measured between Parker and Solar Orbiter, cited in this section. A pair of solid curves of constant source surface longitude shown in both panels isolate a slow (yellow), a fast (blue), and a {\alf}ic slow (red) wind stream observed in both spacecraft. As seen in the bottom panel, this occurs when Parker and Solar Orbiter are also nearly coincident in latitude over an extended range of longitude, which makes this conjunction unique in the mission so far, as we are probing pure radial evolution of multiple streams rather than needing to disentangle radial and latitudinal effects simultaneously.

\subsection{Solar Wind Properties}
\begin{figure*}[]
	\centering
	\includegraphics[width=0.55\linewidth]{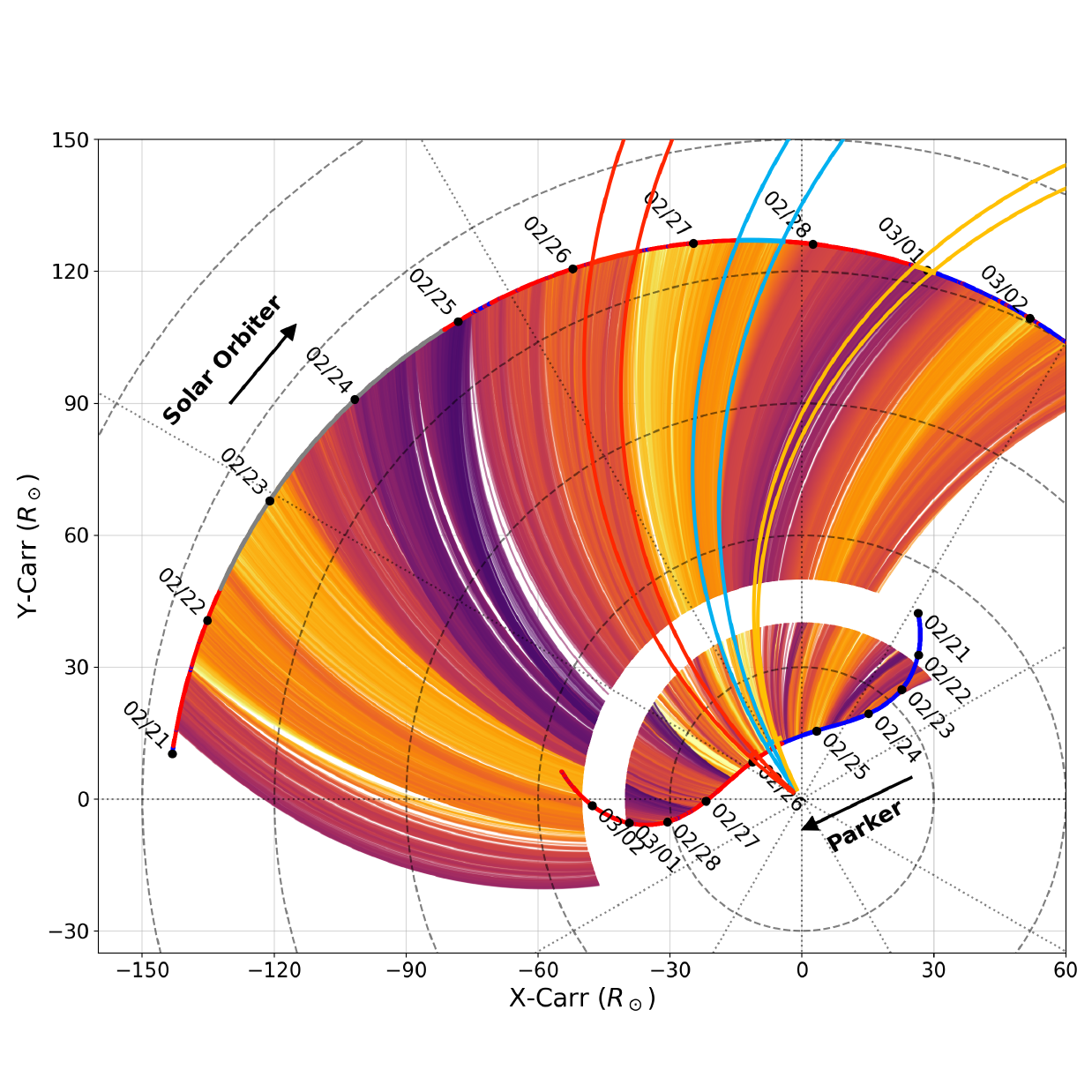}
         \includegraphics[width=0.7\linewidth]{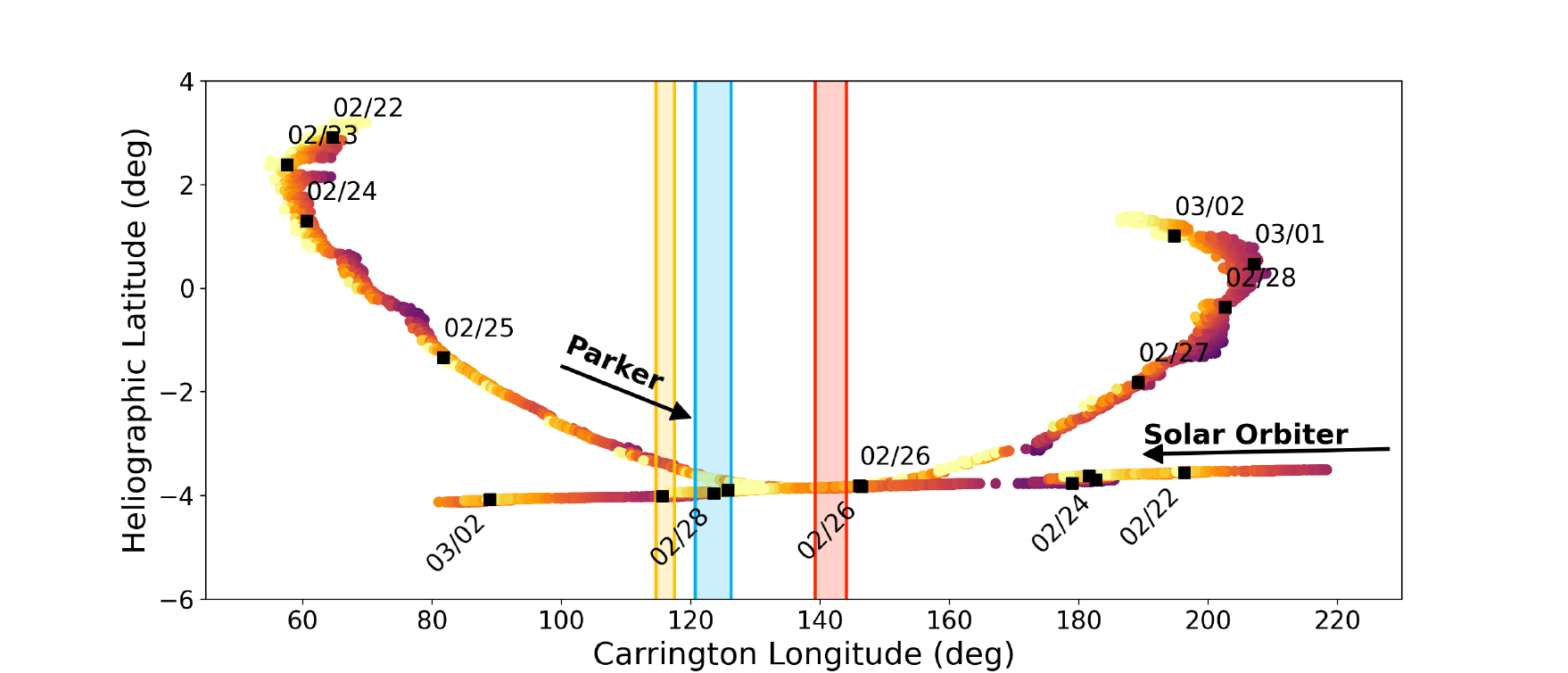}
	\caption{Illustration of the conjunction and ballistic mapping in the Carrington Frame. Top : Solar Orbiter and Parker's trajectories projected into the solar equatorial plane for late February 2022. Colored Parker spiral field lines illustrate how Solar Orbiter and Parker's measurements are ballistically propagated inwards to the Sun.  Bottom: The ballistically mapped heliographic coordinates for Parker and Solar Orbiter at 2.5$R_\odot$. Parker spirals bounded by the red, blue, yellow flow lines (top) and vertical lines (bottom) annotate the slow {\alfic}, fast, and slow stream, respectively, that we follow from Parker out to Solar Orbiter corresponding to the timeframes highlighted in the corresponding colors in Figure \ref{fig:Parker_solo_insitu_data}.}
	\label{fig:relative_pos}
\end{figure*}

\begin{figure*}[]
	\centering
	\includegraphics[width=0.47\linewidth]{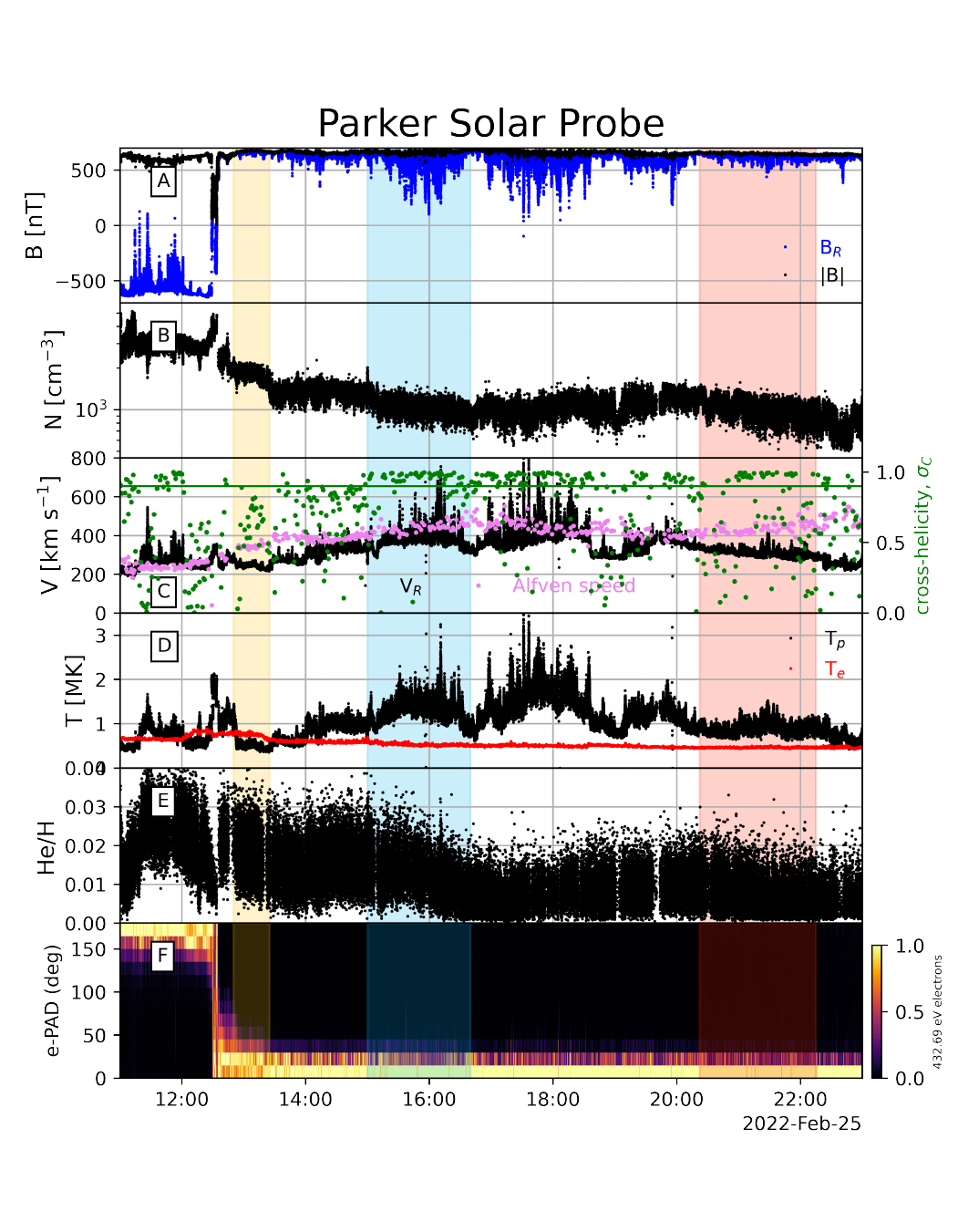}
        \includegraphics[width=0.47\linewidth]{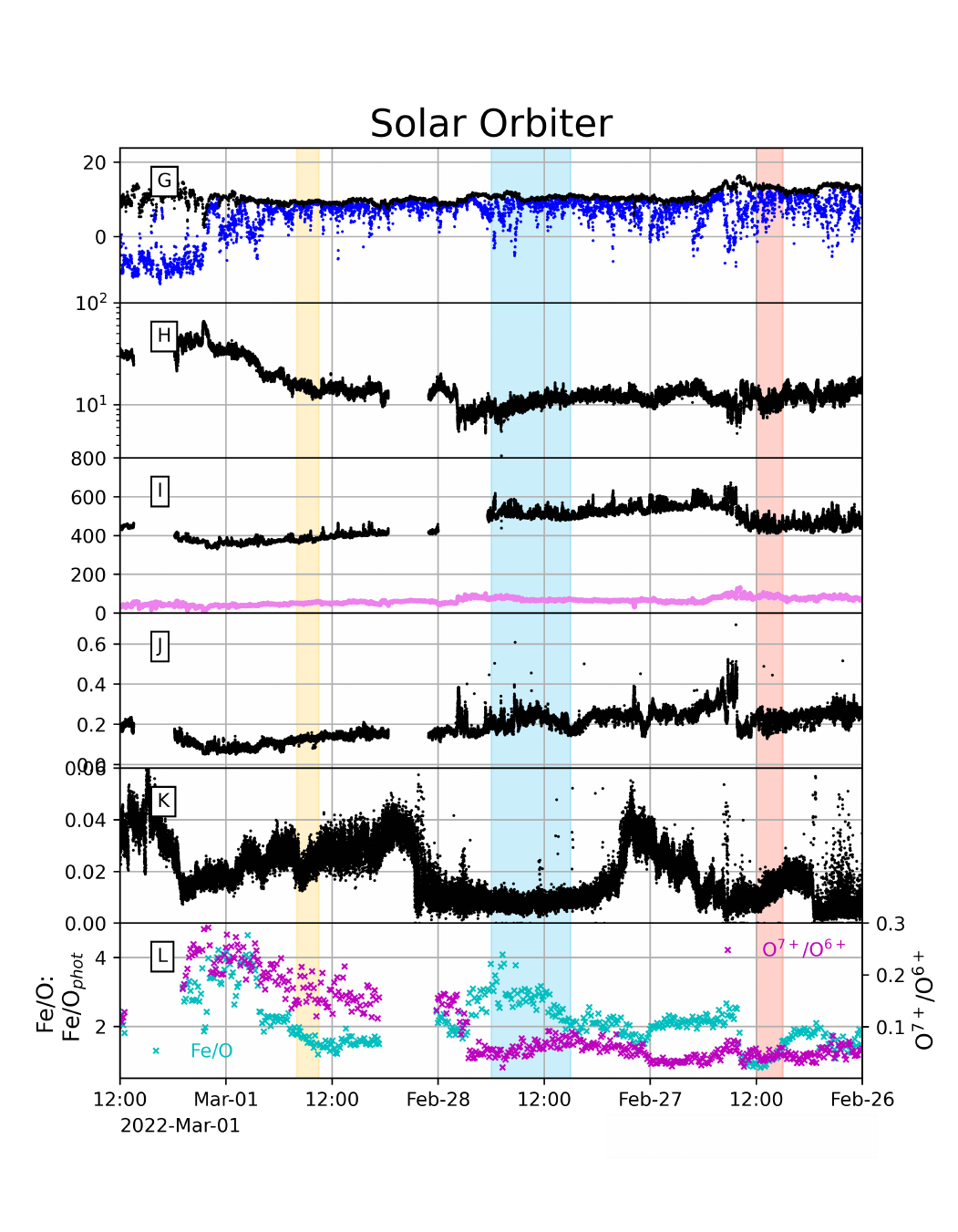}
	\caption{Stackplot showing the solar wind properties at Parker Solar Probe (left) and Solar Orbiter (right), where Solar Orbiter is plotted in reversed time. The shaded region corresponds to the periods of interest. Top to bottom, first five rows are: panel A/G, radial magnetic field component and magnitude, panel B/H, proton density, Panel C/I proton bulk speed, {\alf} speed and cross-helicity (only for Parker) where the horizontal green line indicates a value of 0.9, panel D/J, isotropic proton and electron (only for Parker) temperature, and Panel F/K is the helium abundance as the He/H number density ratio. Panel F is the electron PAD at 432.69 eV. Panel L shows the Fe/O:Fe/O$_{phot}$ and O$^{7+}$/O$^{6+}$ ratio.}
	\label{fig:Parker_solo_insitu_data}
\end{figure*}

Once a connection was established, we identify the slow and {\alf}ic slow  solar wind periods by their asymptotic speed at Solar Orbiter and their {\alficy} as characterized by the normalized cross-helicity at Parker where this signature is most pristine \citep{Matthaeus1982, Bruno2013}. The solar wind properties from this study corresponding to the portion of the streams, backmapped in Figure \ref{fig:relative_pos}, are shown in Figure \ref{fig:Parker_solo_insitu_data} and summarized in Tables \ref{table:CharacteristicPropertiesSlow} and \ref{table:CharacteristicPropertiesAlfvenicSlow}. The figure shows the properties of the solar wind at Parker, left column, and Solar Orbiter, right column, in time. We note that the Solar Orbiter data is plotted in reverse time such that the streams appear left to right. Panel A shows the magnetic field magnitude and radial component, panel B shows the proton density, panel C shows the radial speed (black), {\alf} speed (pink) as $V_{A}=|\textbf{B}|/\sqrt{\mu_{0} m_p n_p}$, and normalized cross-helicity (green) along with a green line noting the 0.9 threshold. Panel D shows the proton (black) and electron (red) temperature. Panel E shows the helium abundance as He/H and panel F the column normalized electron pitch angle distribution (e-PAD) of suprathermal electrons across $0-180^{\circ}$ in the energy range of 432.69eV. Panels G, H, J, K are the same properties at Solar Orbiter. Panel I, we include the proton speed and {\alf} speed and panel L shows the heavy ion composition of the solar wind, showing the Fe$^{6-20+}$/O$^{5-8+}$ normalized to Fe/O$_{phot}\sim2.5$, and $\sim1$, respectively, using a photospheric elemental abundance of Fe/O$_{phot}=0.059$ \citep{Asplund2021}. We also plot the O$^{7+}$/O$^{6+}$ ion ratio.

We compute the cross-helicity as $\sigma_C = 2\delta \textbf{v} \cdot \delta \textbf{b}/(\delta \textbf{v}^{2} + \delta \textbf{b}^{2})$, where the $\delta \textbf{b} = \delta \textbf{B}/|\textbf{B}|\cdot V_{A}$ are in {\alf} units \citep{Barnes1974, Matthaeus1982, Roberts1987}. $\delta B$ is computed as the magnetic field deviation from the mean field, where $\delta \textbf{B} = \textbf{B}-<\textbf{B}>_t$, where $<\textbf{B}>_t$ is time-averaged over 10 minute intervals, $t$, at Parker and Solar Orbiter on the order of typical {\alfic} fluctuations \citep{Tu1995}. Where $\delta v$ is computed in the same manner. The cross-helicity can range between -1 and 1, where a value near $\pm1$ indicates strong correlation/anti-correlation between the magnetic and velocity fluctuations when the stream is {\alfic} \citep{Perri2010, DAmicis2019}. 

We examine a specific portion of the slow and {\alf}ic slow solar wind highlighted in yellow and red, respectively, of the figure. Parker measured the slow speed stream (yellow) with a relatively low cross-helicity (Panel C) between 12:50-13:30UT on 2022 February 25, while measurements from Solar Orbiter of the same stream took place on 13:30-16:00UT on 2022 February 28 with an measured speed of 381 km s$^{-1}$ (Panel I). For the {\alf}ic wind (red), Parker sampled the stream at 15:00-17:00UT on 2022 February 25 with a high cross-helicity (Panel C) and Solar Orbiter at 9:00-12:00UT on 2022 February 26 with a speed of 451 km s$^{-1}$ (Panel I). We also include the fast wind stream examined in \cite{Rivera2024} for comparison. The slow stream is immediately adjacent to a heliospheric current sheet (HCS) crossing measured by both spacecraft as indicated by the change in polarity (Panel A) and electron PAD direction (Panel F). The slow {\alfic} stream is adjacent to several faster and hotter streams seen in both spacecraft. We note that this slow stream measured by Parker begins with a transition from a large proton beam, exemplified by the large temperature increase in panel D, presumably from reconnection within the HCS as described in \citep{Phan2022}.

The periods in the shaded regions of interest are characterized by a number of clear correspondences between the Parker and Solar Orbiter in situ data. The backmapped streams at both spacecraft match in positive field polarities (Panel A and G) and contain persistent A$_{He}\sim\%2$, or He/H$\times100$, and A$_{He}\sim\%1$, for the slow and {\alf}ic slow solar wind shown in panels E and K, respectively. 

Additionally, we measure the Fe/O FIP bias and O$^{7+}$/O$^{6+}$, in panel L, that exhibit some variability across the distinct streams. The mean Fe/O FIP bias for the slow, fast, and {\alfic} slow wind stream is $1.63\pm0.19$, $2.3\pm0.23$, $1.09\pm0.18$, respectively. The variation in the elemental composition of the solar wind suggests distinct sources at the Sun as discussed in Section \ref{sec:coronal_properties} in connection to remote observations. The ion ratio ranges from $0.17\pm0.03$, $0.06\pm0.02$, $0.05\pm0.01$, respectively. The ion ratio is a proxy for the electron temperature at the corona where a higher ratio indicates a hotter source electron temperature. Solar wind around the HCS and in the slowest speed wind is generally more highly ionized compared to fast speed wind, in line with the present observations \citep{Lepri2013, Xu_Borovsky2015}.

The chosen source surface longitude ranges and associated timeframes yield mass and magnetic flux conservation between the spacecraft, as discussed in Section \ref{sec:energyconservation}. We note that the {\alf}ic slow stream (red) contains a relatively lower amplitude switchback patch in the Parker data as compared to the fast wind (blue), which are nearly at the same heliocentric distance. While  both streams are sub-{\alfic} as shown in Panel C, the {\alfic} slow wind shows a lower mach number, V$_{SW}$/V$_{A}\sim0.71$, compared to the fast wind case ($\sim 0.89$) which has been connected to larger deflection angles in the magnetic field that may be connected to their lower B$_R$ amplitude \citep{Liu2023}. The switchback patch is identified as having correlated, large amplitude, B$_r$ and v$_R$ fluctuations while maintaining near-constant magnetic field magnitude and persistent electron strahl direction, as shown in panels A, C, F, respectively. Both B$_R$ and v$_R$ range in amplitude at Parker (between different patches) while appearing overall less coherent in the Solar Orbiter observations. 

The fast and {\alf}ic slow solar wind related to switchback patches occur at markedly different speeds. At Parker, the fast wind stream has an average bulk speed of 386 km s$^{-1}$ with a speed of 512 km s$^{-1}$ at Solar Orbiter. For the {\alf}ic slow wind, the mean speed at Parker is 311 {\kms} with acceleration to 451 km s$^{-1}$ at Solar Orbiter. In both cases, the solar wind undergoes significant acceleration between the spacecraft separation, $\sim$13.3--130 $R_\odot$, while also suggesting that the acceleration experienced by individual solar wind streams is related to the amplitude of the switchbacks near the Sun.

\begin{table}
\centering
\begin{tabular}{c c c} \hline
Properties \\ Slow Wind & Parker Solar Probe  & Solar Orbiter \\
& [$13.35R_{\Sun}$] & [$125R_{\Sun}$] \\ \hline \hline

n$_p$ [cm$^{-3}$] & $1712\pm244$ & $14.8\pm2.4 $ \\ \hline
v$_{p}$ [km s$^{-1}$] & $251\pm16 $ & $381\pm10 $ \\ \hline
T$_{p}$ [MK] & $0.6\pm0.3 $ & $ 0.12\pm0.1 $ \\ \hline
V$_{A}$ [km s$^{-1}$] & $ 351\pm22 $ & $53\pm13 $ \\ \hline
B$_r$ [nT] & $ 658\pm11 $ & $8\pm2 $ \\ \hline
$\vert$B$\vert$ [nT] & $ 667\pm12 $ & $ 9\pm0.3 $ \\ \hline

n$_{\alpha}$/n$_p$ & $0.02\pm0.005 $ & $0.02\pm0.004 $ \\ \hline
v$_{\alpha}$ [km s$^{-1}$] & $ 336\pm36 $ & $ 449\pm6 $ \\ \hline
T$_{\alpha}$ [MK] & $4.6\pm2.6$ & $0.65\pm0.07$ \\ \hline
T$_{e}$ [MK] & $ 0.75\pm0.01$ & $ 0.27\pm0.01$\footnote{In lieu of a direct measured constraint, we estimate $T_e|_{Solar Orbiter}$ using $T_e|_{Parker}$ and the statistical electron polytropic index derived by \cite{Dakeyo2022} for their solar wind family C.} \\ \hline
$\delta$B [nT] & $90\pm46 $ & $5.3\pm2.6 $ \\ \hline

\end{tabular}\\
\caption{Characteristic Parker and Solar Orbiter properties for the slow wind averaged within the yellow regions of Figure \ref{fig:Parker_solo_insitu_data}.}
\label{table:CharacteristicPropertiesSlow}
\end{table}

\begin{table}
\centering
\begin{tabular}{c c c} \hline
Properties \\ {\alfic} Slow Wind & Parker Solar Probe  & Solar Orbiter \\
& [$13.7R_{\Sun}$] & [$130.6R_{\Sun}$] \\ \hline \hline

n$_p$ [cm$^{-3}$] & $1043\pm211 $ & $11\pm2.8$ \\ \hline
v$_{p}$ [km s$^{-1}$] & $311\pm22 $ & $451\pm21 $ \\ \hline
T$_{p}$ [MK] & $0.9\pm0.12 $ & $ 0.22\pm 0.03$ \\ \hline
V$_{A}$ [km s$^{-1}$] & $ 439\pm15 $ & $88\pm7 $ \\ \hline
B$_r$ [nT] & $ 631\pm21 $ & $9.4\pm3 $ \\ \hline
$\vert$B$\vert$ [nT] & $ 650\pm7 $ & $13.3\pm0.34 $ \\ \hline

n$_{\alpha}$/n$_p$ & $0.01\pm0.005 $ & $0.016\pm0.007 $ \\ \hline
v$_{\alpha}$ [km s$^{-1}$] & $434\pm53 $ & $537\pm14 $ \\ \hline
T$_{\alpha}$ [MK] & $6.6\pm2.4$ & $1.1\pm0.15 $ \\ \hline
T$_{e}$ [MK] & $0.47\pm0.01$ & $0.17\pm0.01$\footnote{In lieu of a direct measured constraint, we estimate $T_e|_{Solar Orbiter}$ using $T_e|_{Parker}$ and the statistical electron polytropic index derived by \cite{Dakeyo2022} for their solar wind family D.} \\ \hline
$\delta$B [nT] & $140\pm71 $ & $13\pm4.6 $ \\ \hline

\end{tabular}\\
\caption{Characteristic Parker and Solar Orbiter properties for the {\alf}ic slow wind averaged within the red regions of Figure \ref{fig:Parker_solo_insitu_data}.}
\label{table:CharacteristicPropertiesAlfvenicSlow}
\end{table}

\begin{figure*}[]
	\centering
	\includegraphics[width=0.45\linewidth]{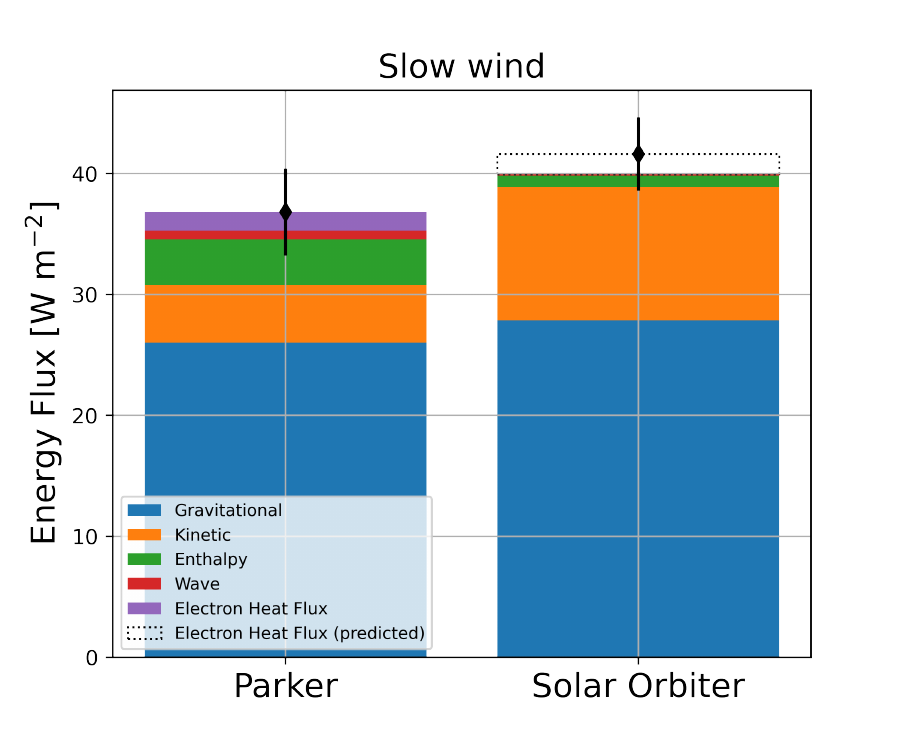}
        \includegraphics[width=0.45\linewidth]{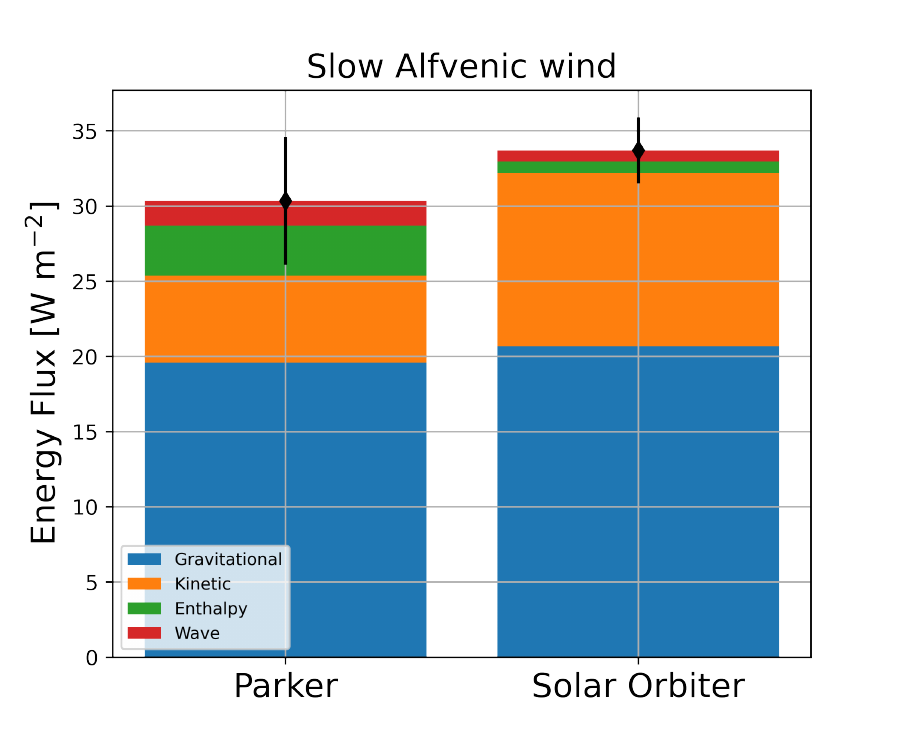}
	\caption{Energy fluxes at Parker and Solar Orbiter for the slow (left) and {\alf}ic slow (right) solar wind. The values are computed from Equation \ref{eq:Energyconservation} and listed in Table \ref{table:CharacteristicPropertiesSlow} and \ref{table:CharacteristicPropertiesAlfvenicSlow}}.
	\label{fig:energy_fluxes}
\end{figure*}

\begin{figure*}[]
	\centering
 \includegraphics[width=0.49\linewidth]{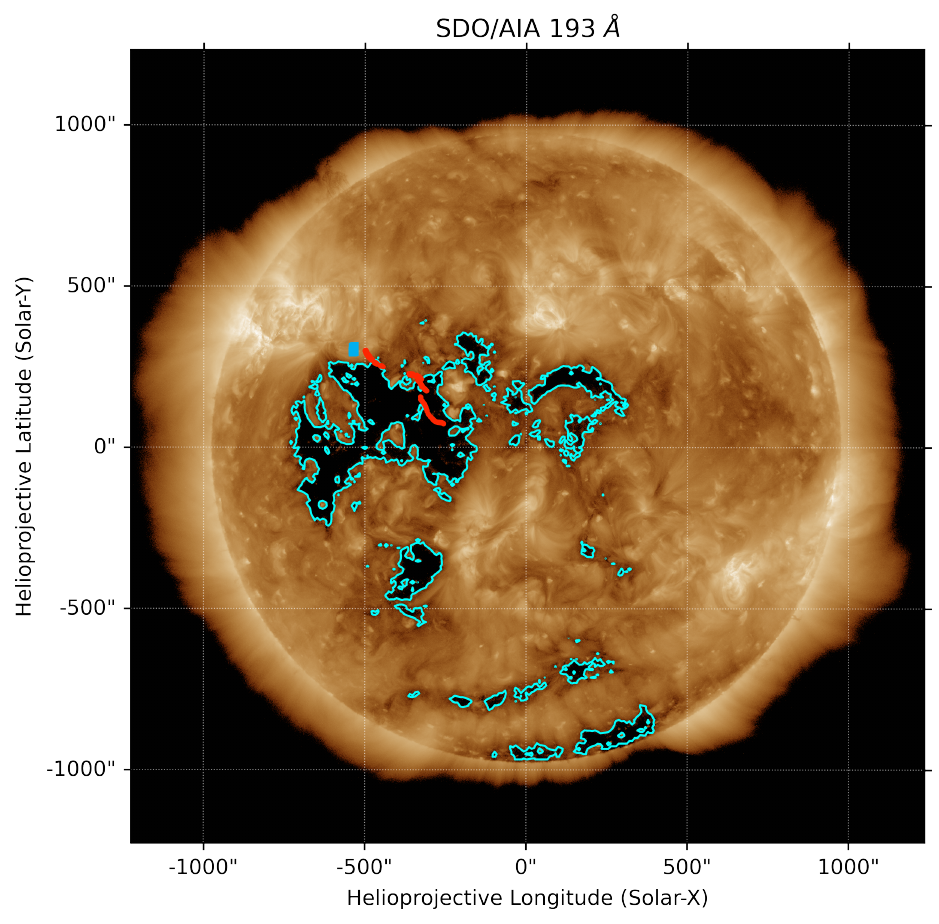}
	\includegraphics[width=0.49\linewidth]{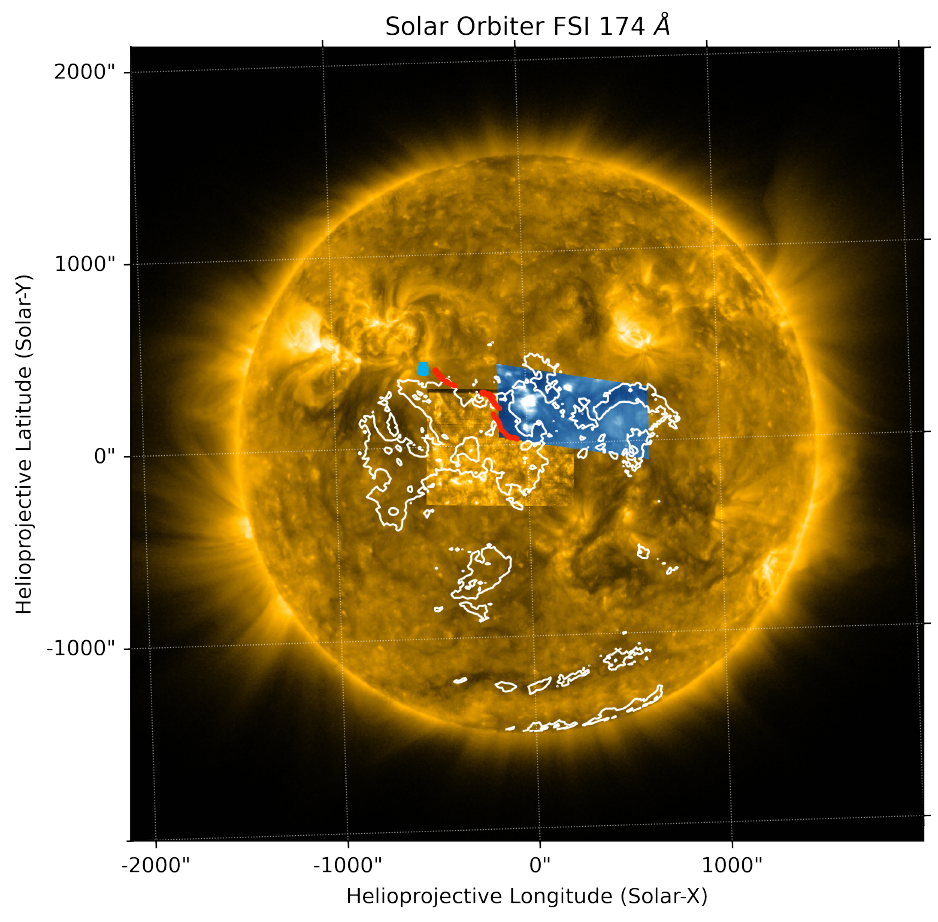}
	\caption{Full disk images of SDO/AIA 193\AA\;on 2022-02-23 00:30UT (left) and Solar Orbiter FSI 174\AA\;(right) on 2022-02-23 11:21UT. The outline indicates the coronal hole boundary determined by AIA 193\AA~and mapped to FSI 174\AA. The FSI image contains the SPICE raster of the Ne VIII 770.42\AA\;taken on 2022-02-23 11:23-10:06UT and Hinode/EIS raster of the Fe XII 192.813\AA\;taken on 2022-02-23 00:30-01:31UT.}
	\label{fig:solo_remote_data_fulldisk}
\end{figure*}

\begin{figure*}[]
	\centering
	\includegraphics[width=0.45\linewidth]{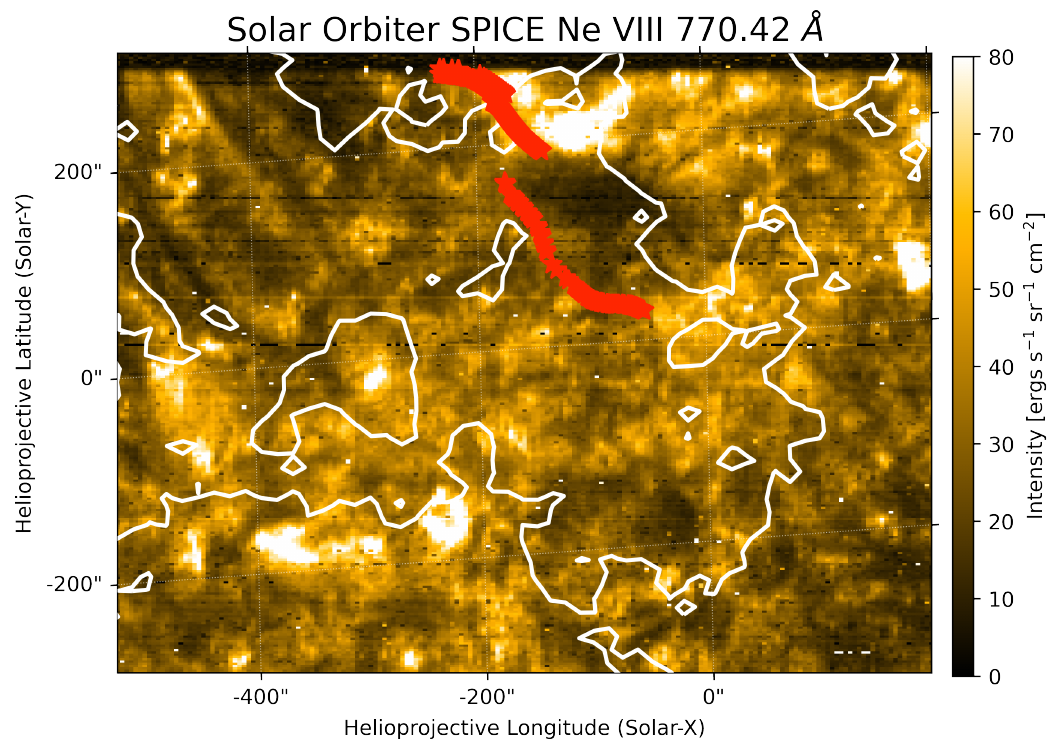}
    \includegraphics[width=0.45\linewidth]{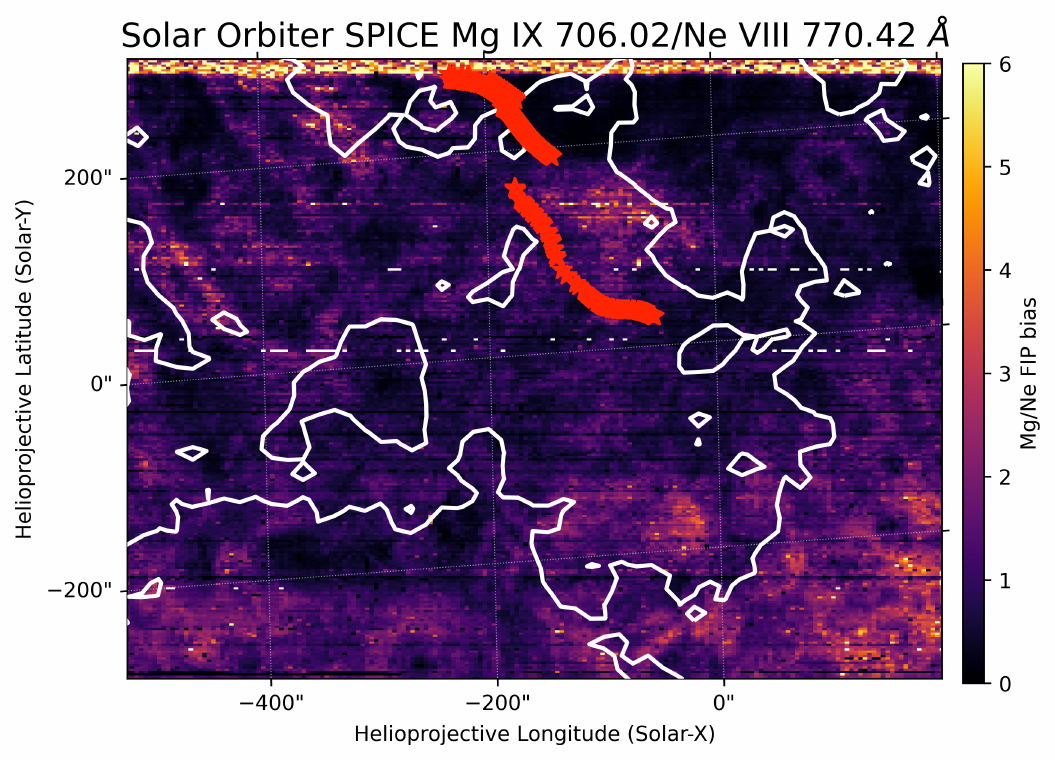}
	\caption{Solar Orbiter SPICE intensity of the Ne VIII 770.42 \AA~ (left) and FIP bias of Mg/Ne (right). The white outline is the coronal hole outline from Figure \ref{fig:solo_remote_data_fulldisk}. The red stars are the footpoint mapping for the {\alf}ic slow solar wind corresponding to the red timeframe in Figure \ref{fig:Parker_solo_insitu_data}.}
	\label{fig:solo_remote_data_SPICE}
\end{figure*}

\begin{figure*}[]
	\centering
	\includegraphics[width=0.45\linewidth]{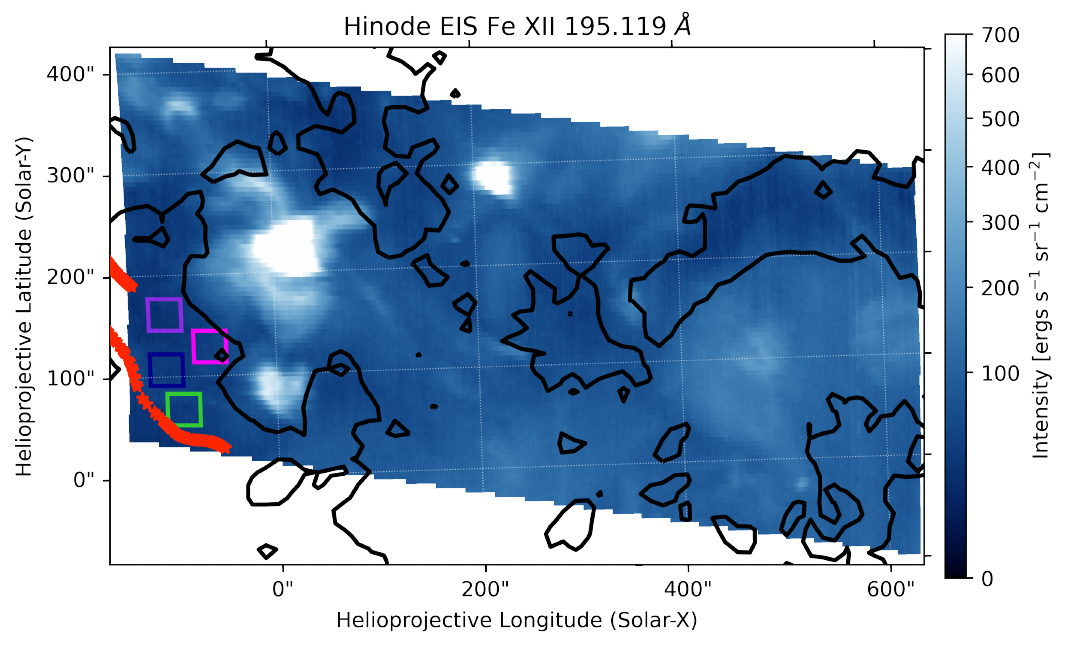}
 \includegraphics[width=0.45\linewidth]{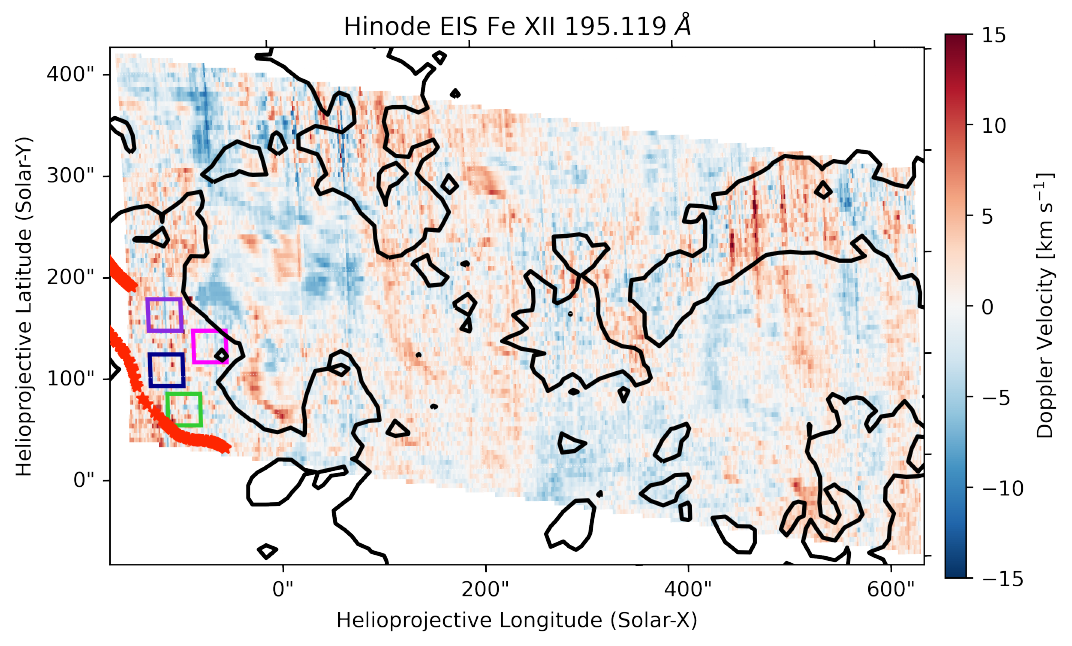}
	\caption{Hinode/EIS observations of the Fe XII 195.119 \AA~fitted intensity (left) and velocity Doppler shifts (right). The white (left) and white (right) outline is the coronal hole outline from Figure \ref{fig:solo_remote_data_fulldisk}.}
	\label{fig:EIS_rmote_data}
\end{figure*}

\begin{figure}[]
	\centering
	\includegraphics[width=\linewidth]{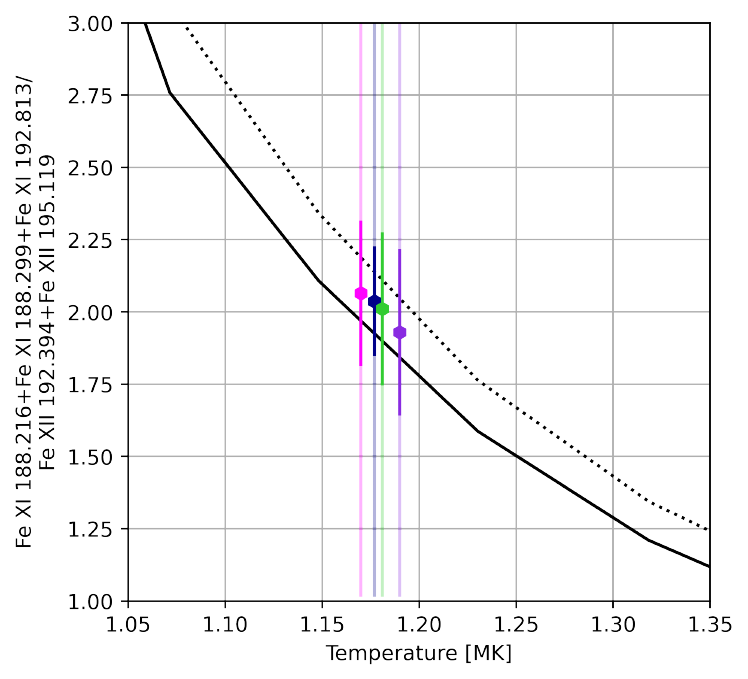}
	\caption{Ratio of the average intensity within each box in Figure \ref{fig:EIS_rmote_data} for the sum of the Fe {\footnotesize XI} 188.216 + Fe {\footnotesize XI} 188.299 + Fe {\footnotesize XI} 192.813 to the sum of Fe {\footnotesize XII} 195.119 + Fe {\footnotesize XII} 192.394 shown for each corresponding color. The theoretical curve of the intensity ratio across electron temperature are in black, as derived using CHIANTI v.10 atomic values for the minimum (solid) and maximum (dashed) electron density computed from the Fe XIII line ratio.}.
	\label{fig:EIS_temps}
\end{figure}


\begin{table*}
\centering
\begin{tabular}{ccccc}
	\hline
	
Slow Wind: Energy Flux Terms $\times$(r/R$_{\odot})^{2}$ & Parker [W m$^{-2}$] & Solar Orbiter [W m$^{-2}\times f$] & $\Delta$ Solar Orbiter -- Parker \\ \hline \hline

$W_{Kinetic}$: & $4.76\pm0.68$ & $11.02\pm0.95$ & $6.26\pm1.17$ \\ \hline 

$W_{Enthalpy}$:  & $3.76\pm0.43$ & $0.93\pm0.08$ &  $-2.83\pm0.44$\\ \hline

$W_{Gravitational}$: & $26.03\pm2.08$ & $27.87\pm2.01$ & $1.84\pm2.89$ \\ \hline

$W_{Wave}$: & $0.73\pm0.58$ & $0.13\pm0.11$ & $-0.60\pm0.59$\\ \hline

$W_{Qe}$: & $1.54\pm0.51$ & $1.61\pm0.53^{\dagger}$ & $0.07\pm0.74$\\ \hline

Total & $ 36.82\pm3.60$  & $ 41.56\pm3.04$ & $4.74\pm4.30$\\ \hline

\end{tabular} \\
\caption{Individual energy terms and standard deviation computed from Equation \ref{eq:Energyconservation} for Parker and Solar Orbiter averaged across source surface longitude $116-118^{\circ}$ (as indicated in Figure \ref{fig:relative_pos}) and normalized by the flow speed and solar radii. The values of Solar Orbiter are multiplied by a characteristic expansion factor, $f_{slow}$, computed from mass conservation in Equation \ref{eq:massconserveslow}. $^{\dagger}$Calculated based on an equal relative energy flux at Solar Orbiter.}
\label{table:EnergyTermsslow}
\end{table*}

\begin{table*}
\centering
\begin{tabular}{ccccc}
	\hline
	
{\alfic} Slow Wind Energy: Flux Terms $\times$(r/R$_{\odot})^{2}$ & Parker [W m$^{-2}$] & Solar Orbiter [W m$^{-2}\times f$] & $\Delta$ Solar Orbiter -- Parker \\ \hline \hline

$W_{Kinetic}$:  & $5.78\pm1.53$ & $11.54\pm2.31$ & $5.76\pm2.77$ \\ \hline 

$W_{Enthalpy}$: & $3.10\pm0.70$ & $0.76\pm0.15$ &  $-2.34\pm0.72$\\ \hline

$W_{Gravitational}$: & $19.59\pm2.95$ & $20.67\pm2.80$ & $1.08\pm4.06$ \\ \hline

$W_{Wave}$: & $1.64\pm1.1$ & $0.73\pm0.45$ & $-0.91\pm1.19$\\ \hline

Total & $30.34\pm4.24$  & $33.70\pm2.20$ & $3.36\pm4.78$\\ \hline

\end{tabular} \\
\caption{Individual energy terms and standard deviation computed from Equation \ref{eq:Energyconservation} for Parker and Solar Orbiter averaged across source surface longitude $140-146^{\circ}$ (as indicated in Figure \ref{fig:relative_pos}) and normalized by the flow speed and solar radii. The values of Solar Orbiter are multiplied by a characteristic expansion factor, $f_{alfvenic}$, computed from mass conservation in Equation \ref{eq:massconservealfvenic}.}
\label{table:EnergyTermsalfvenic}
\end{table*}

\section{Conservation Equations in the Flux Tube} \label{sec:conservationequations}

\subsection{Mass and Magnetic Flux Conservation}
Before we address the conservation of energy, however, we must first establish whether the system is acting like a 1-dimensional flux tube. We introduce the solid angle subtended by the flux tube at each spacecraft ($\Omega$) and refer to the ratio of this at both spacecraft as the streams expansion factor $f$. We expect each stream to conserve mass flux such that,

\begin{align}
    \frac{dM}{dt}\bigg|_{Parker}^{Solar~Orbiter} = \Omega n_p u_{cm} r^{2}\Big|_{Parker}^{Solar~Orbiter} = 0
\end{align}

Using the characteristic values in Table \ref{table:CharacteristicPropertiesSlow} and \ref{table:CharacteristicPropertiesAlfvenicSlow}, we compute that mass conservation for the slow solar wind stream occurs for expansion factor:

\begin{eqnarray}
\nonumber
    f_{slow} = \frac{\Omega_{Solar Orbiter}}{\Omega_{Parker}} =  \frac{n_p u_{cm}r^2\big|_{Parker}}{n_pu_{cm}r^2\big|_{Solar Orbiter}} \\
        = 0.91 \pm 0.13 
       \label{eq:massconserveslow}
\end{eqnarray}
and for the {\alf}ic slow solar wind,
\begin{eqnarray}
\nonumber
    f_{alfvenic} = \frac{\Omega_{Solar Orbiter}}{\Omega_{Parker}} =  \frac{n_p u_{cm}r^2\big|_{Parker}}{n_pu_{cm}r^2\big|_{Solar Orbiter}} \\ 
        = 0.72 \pm 0.21
       \label{eq:massconservealfvenic}
\end{eqnarray}

The $\Omega$ is the angular wedge traversed across the flux tube in the individual streams. The center of mass velocity is computed as,

\begin{equation}
u_{cm} = \frac{\Sigma_{j} m_{j}n_{j} v_{j}}{\Sigma_{j} m_{j} n_{j}}.
\end{equation}

$r$ is the heliocentric distance of the spacecraft and $n_{j}$, $m_{j}$, and $v_{j}$ for the species, $j\in[p,\alpha]$, that are the protons and alpha particles. Therefore the tube is approximately conical ($f$ is near unity in each case) but has under expanded by a factor of up to $\sim13\%$ and 28\% for the slow and {\alf}ic slow cases, respectively. We therefore set the flux tube expansion factor, $f= 0.91$ and  0.72 between the spacecraft. \cite{Rivera2024} finds the flux tube expansion factor for the fast solar wind period (blue shaded region in Figure \ref{fig:Parker_solo_insitu_data}) in this encounter to be $0.90\pm0.09$. This is similar to the analysis in a $\sim 600$ km s$^{-1}$ coronal hole stream from \cite{Schwartz1983} which finds a value of $f \sim 0.88$ for a Helios 1 and 2 spacecraft lineup (0.2AU separation) suggesting some flux tube compression between 0.5 to 0.72AU in line with the present study. 

To confirm this underexpansion in terms of mass flux is balanced by a corresponding change in magnetic flux, we also assess conservation of mass and magnetic flux together across our flux tube. Since this environment is confined to a single positive polarity magnetic sector, we do not expect any annihilation, although some reconnection has been reported in the form of jets beyond 0.6au in slower speed solar wind \citep{Fargette2023}. We can check simply by relating the magnetic flux scaling of $B_r\cdot r^2$ to the mass flux n$_p\cdot$v$_p\cdot$r$^{2}$, as n$\cdot$v$_p$/B$_r$ = constant. Using the average values the timeframes of interest from Table \ref{table:CharacteristicPropertiesSlow} and \ref{table:CharacteristicPropertiesAlfvenicSlow}, we compute a ratio of $1.10\pm0.11$ and $1.05\pm0.16$, respectively. A near-unity value of this quantity indicates that the magnetic and mass fluxes both indicate the same expansion factor for the stream, and the streams therefore behave as conservative 1D flux tubes between the two spacecraft. 

\subsection{Conservation of Energy} \label{sec:energyconservation}
In order to model the conservation of energy in the solar wind stream, we model it as a persistent, radial flux tube of solid angle $\Omega$. In the observations, the spacecraft traverses across a flux tube $\sim 2$ degrees wide for the slow wind and $\sim 6$ degrees wide for the {\alfic} slow wind. The plasma is sampled at distances centered on approximately $r_{Parker} \sim 13.4R_{\Sun}$ and $r_{Solar~Orbiter} \sim 124R_{\Sun}$ for the slow wind and $r_{Parker} \sim 13.8R_{\Sun}$ and $r_{Solar~Orbiter} \sim 131R_{\Sun}$ for the {\alf}ic slow wind. Physical quantities are assumed to be uniform across the cross section of each flux tube and therefore to only vary with heliocentric distance.

In both streams, significant contributions come from the kinetic energy of the plasma bulk flow, $n_jm_jv_j^2/2$, the thermal energy carried by the plasma, $n_jk_B T_j$, and gravitational energy, $m_jn_jGM_{\Sun}(1-R_{\Sun}/r)/R_{\Sun}$, where the species which we consider, $j\in[e,p,\alpha]$, are the protons, electrons and alpha particles. As the slow and {\alf}ic slow wind includes non-compressible {\alf}ic fluctuations, we also consider the energy associated with these waves, $\delta B^2/2\mu_0$, carried by the {\alf} waves throughout these periods. We also quantify the associated heat fluxes which can become non-negligible in the slowest wind streams \citep{Halekas2023}.

We can write the conservation of energy flux, $W$, in the flux tube segment in terms in the center of mass frame of these components following the analysis from \cite{Jacques1977, Rivera2024}:
\begin{eqnarray}
    W\bigg|_{Parker}^{Solar~Orbiter} = u_{cm} r^2/R_{\Sun}^2 \Omega \nonumber \\ \bigg( \frac{n_{j} m_{j} u_{cm}^2}{2}+ \frac{5}{2}n_{j}k_{B}T_{j} \nonumber \\
     + \frac{m_{j}n_{j}GM_{\Sun}}{R_{\Sun}} (1-R_{\Sun}/r) + \nonumber \\
   (3/2+V_A/u_{cm})\frac{\delta B^2}{2\mu_0} + E_{Qe} \bigg) \Big|_{Parker}^{Solar ~Orbiter} = 0.
        \label{eq:Energyconservation}
\end{eqnarray}
Therefore, the system is in a steady state if the change to the energy flux is zero. The first four components include the kinetic, enthalpy, gravitational, and wave energy terms where the leading term outside the parenthesis is a normalizing factor, $r^2/R_{\Sun}$, scaled to a solar radius \citep{Liu2021}. We also compute the electron (E$_{Qe}$) and proton heat fluxes for the two periods at Parker. We assume a polytropic index for an ideal gas with 3 degrees of freedom ($\gamma=5/3$) to compute the enthalpy term. The contribution of the electron heat flux in the {\alf}ic slow wind period is small, being on the order of 0.01 W m$^{-2}$ (or 0.02\%), while several orders of magnitude larger for the slow wind period, at $\sim4\%$. Therefore, we include the contribution of the electron heat flux, as calculated in \citep{Halekas2023}, in the energy conservation of the slow wind period. However, we note that the electron heat flux was only computed at Parker because of the unavailable electron measurements at Solar Orbiter for this timeframe and assumed to contribute the same energy flux percentage at Solar Orbiter, consistent with statistical results from \citep{Halekas2023}. Therefore, the overall relative contribution to the total energy flux is the same at Parker and Solar Orbiter, and that energy is conserved. We calculate the proton heat flux to be $<<1\%$ in each stream and therefore exclude it from the energy flux calculation.

In our equation, $n_{j}$, $m_{j}$, $T_{j}$ are the number density, mass, and temperature, respectively, of the protons, electrons, and alpha particles. $B$ is the magnetic field and V$_{A}$ is the local {\alf} speed. $G$, $k_B$, $M_{\Sun}$, $R_\odot$, $\mu_0$, is the gravitational constant, Boltzmann constant, mass and radius of the Sun, and permeability of free space, respectively. 

Tables \ref{table:EnergyTermsslow} and \ref{table:EnergyTermsalfvenic} lists the averages for individual energy terms calculated in the center of mass frame for the total of the proton, alpha, and electron populations following Equation \ref{eq:Energyconservation} to determine if the energy supplied at the base of the flux tube is conserved between the spacecraft. The Solar Orbiter energy fluxes at each stream are multiplied by their individual expansion factor, $f_{slow}=0.9$ and $f_{alfvenic}=0.72$, to appropriately account for their compression. Figure \ref{fig:energy_fluxes} is a visual representation of the contribution of the individual energy flux terms at Parker and Solar Orbiter, multiplied by their compression factors, for the slow wind and {\alf} slow wind. The figures include all the energy terms listed in Tables \ref{table:EnergyTermsslow} and \ref{table:EnergyTermsalfvenic}. As the figure illustrates, and as indicated by the total energy flux in the related tables, energy conservation is met within the standard deviation of the total energy flux for both solar wind streams. 

Under the well-supported hypothesis that these are the same streams, as indicated through the conserved quantities and stream properties discussed, we explore the degree to which dissipation of {\alf} waves in each stream may be the source of sufficient energy to allow for a self-consistent (i.e. mass and energy conserving) heating and acceleration.

\begin{table*}
\centering
\begin{tabular}{c c c c c } \hline
Spectral Line & Intensity [Purple] & Intensity [Blue] & Intensity [Green] & Intensity [Magenta] \\ \hline \hline
       Fe {\footnotesize XI} 188.216 & 67 & 62 & 66 & 71\\ \hline
       Fe {\footnotesize XI} 188.299  & 59 & 54 & 59 & 63 \\ \hline
       Fe {\footnotesize XI} 192.813  & 49 &  45 & 49 & 52\\ \hline
       Fe {\footnotesize XII} 192.394  & 21 & 19 & 22 & 23\\ \hline
       Fe {\footnotesize XII} 195.119 & 70 & 60 & 65 & 69 \\ \hline
       Fe {\footnotesize XIII} 202.044 & 82 & 82 & 49 & 58 \\ \hline
       Fe {\footnotesize XIII} 203.826 & 27 & 13 & 21 & 49 \\ \hline
\end{tabular}

\caption{List of average intensities for different spectral lines of each box shown in Figure \ref{fig:EIS_rmote_data} that are used in the temperature calculation in Figure \ref{fig:EIS_temps}. Intensities are in units of ergs s$^{-1}$ sr$^{-1}$ cm$^{-2}$ and spectral lines are in \AA.}

\label{table:EIS_intensities}
\end{table*}

\section{Coronal Properties} \label{sec:coronal_properties}
\subsection{Solar Wind Source Mapping} \label{sec:mapping}
To explore the coronal properties of the solar wind and explore its conditions below Parker's orbit, we examine several remote observations from Hinode and Solar Orbiter. The backmapped solar wind trajectories at $2.5 R_\odot$ are subsequently mapped to estimates of the photospheric source locations using Potential Field Source Surface modeling of the coronal magnetic field. The resulting footpoints are overlaid on various remote sensing observations in Figures \ref{fig:solo_remote_data_fulldisk}, \ref{fig:solo_remote_data_SPICE} and \ref{fig:EIS_rmote_data}. 

Figure \ref{fig:solo_remote_data_fulldisk} shows the Sun from the field of view of SDO/AIA in the 193\AA~(left) and Solar Orbiter FSI 174\AA~(right) with the associated fast (blue) and {\alf}ic slow (red) footpoints. The FSI image includes a SPICE Ne {\footnotesize VIII} 770.42 and EIS Fe {\footnotesize XII} 192.813 \AA\/ intensity raster taken the same day. The cyan outline in the AIA 193\AA\/ images indicates the overall boundary of the coronal hole structure determined from an intensity threshold of emission with intensity below 90 DN/s simply as a reference between the remote observations. The coronal hole outline determined from AIA is mapped directly to the FSI 174\AA\/ image as the white outline, as well as the SPICE and EIS observations discussed below. We note the footpoints for the slow wind appear to the left of the fast wind footpoints but are not included in the figures. We find the footpoints associated with the fast and {\alf}ic slow wind are associated with the boundary of the extended equatorial coronal hole network. The footpoints for the {\alf}ic slow solar wind fall within the SPICE raster and at the lower left of the EIS FOV that both observe part of the same coronal hole structure nearly contemporaneously. The SPICE observations provide a comparison between the elemental composition with HIS in situ while the EIS observations provide electron density and temperature constraints in the corona.

\subsection{Elemental Composition}
We compare the remotely observed and in situ elemental abundances from Solar Orbiter to confirm the heliospheric to photosphere mapping is reasonable and to justify the coronal constraint in the solar wind modeling from the source conditions, i.e. electron temperature and density from Section \ref{sec:EIStemp}. Figure \ref{fig:solo_remote_data_SPICE} shows the SPICE raster from Figure \ref{fig:solo_remote_data_fulldisk} showing the intensity (left) and the First Ionization Potential (FIP) bias of Mg/Ne (right) with the same coronal hole outline. The Mg/Ne elemental abundance can be used to compare with the elemental abundances (Fe/O) measured in situ with Solar Orbiter where similarly low-FIP enhanced ratios can be utilized to confirm footpoint connection. The intensity from SPICE is computed by integrating over the fitted Gaussian profile of the Ne and Mg spectral lines. The FIP bias is determined by taking an intensity ratio of Mg {\footnotesize IX} 706.02 \AA (low FIP, 7.6eV) and Ne {\footnotesize VIII} 770.42 \AA (high FIP, 21.6eV) following analysis for SPICE observations in \cite{Brooks2022, Varesano2024}. The intensity governed by collisional excitation in units of phot cm$^{-2}$ s$^{-1}$ arcsec$^{-2}$ is given as,

\begin{eqnarray}
	I_{coll}  =  \frac{1}{4\pi} \int_{-\infty}^{\infty} G(T,n_{e}) \varphi(T) dT  \label{intensity}
\end{eqnarray}
\noindent where $T$ and $n_{e}$ are the electron temperature and density, respectively. $G(T,n_{e})$ is the contribution function in units cm$^{3}$ s$^{-1}$. $\varphi (T)$ is the Differential Emission Measure (DEM) in units of cm$^{-5}$ K$^{-1}$. They are defined as follows:
\begin{eqnarray}
	& G(T,n_{e}) = \frac{n_{j} (X^{+q})}{n(X^{+q})} \frac{n(X^{+q})}{n(X)} \frac{n(X)}{n(H)} \frac{n(H)}{n_{e}} \frac{A_{ji}}{n_{e}} \label{contribution} \\
	& \varphi (T) = n_{e}^{2}\frac{dx}{dT} \label{DEM}
\end{eqnarray}
\noindent where $n_{j} (X^{+q})/n(X^{+q})$ is the population of level j of the +q ion of element X, ${n(X^{+q})}/{n(X)}$ is the relative abundance of the +q state, ${n(X)}/{n(H)}$ is the absolute abundance, ${n(H)}/{n_{e}}$ is the Hydrogen to electron ratio and  $A_{ij}$ is the Einstein coefficient for spontaneous emission. The spectral line analysis uses atomic properties and tools from CHIANTI v.10 \citep{delzanna2021} using version 0.2.3 \citep{Barnes2024} of the fiasco open source software package. For the SPICE observations, the spectral lines were chosen because their ions spans a similar temperature range, therefore their ratio is largely independent of temperature. The FIP bias is a ratio of the elemental abundances which can be calculated by separating the elemental abundances from the contribution function, Ab(Mg) $={n(Mg)}/{n(H)}$ to Ab(Ne) $={n(Ne)}/{n(H)}$, and inverting the equation such that,
\begin{eqnarray}
	\frac{Ab_{Mg}}{Ab_{Ne}}  =  \frac{I_{Mg}}{I_{Ne}} \frac{C(T,n_{e})_{Ne}}{C(T,n_{e})_{Mg}}  \label{Abunratio}
\end{eqnarray}
where,
\begin{eqnarray}
	& G(T,n_{e}) = Ab\times C(T,n_{e}). \label{newcontribution}
\end{eqnarray}

Under ionization equilibrium conditions, the peak of the formation temperature curve for Mg {\footnotesize IX} is log$_{10}$(T [K]) = 6.0 and Ne {\footnotesize VIII} is log$_{10}$(T [K]) = 5.8. Therefore, the emission observed is largely associated with plasma with temperatures spanning typical transition region and low corona plasma associated in the quiet Sun and coronal holes.

We note there is a non-negligible difference in peak formation temperature in the two lines chosen for the abundance analysis. Therefore, changes in the Mg/Ne FIP bias map of Figure \ref{fig:solo_remote_data_SPICE} may arise simply from changes in temperature given that the lines span a slightly shifted formation temperature range. However, it is generally found that coronal holes have small temperature variation and are observed to be fairly stable, isothermal structures in the corona \citep{Hedge2024}. Statistical coronal hole DEMs indicate a narrow peak at $\sim1$MK and a much smaller peak at $\sim1.5$ that remain consistent across several solar rotations \citep{Saqri2020}. However, they may have some small radial variation \citep{Landi2008}. The variation in the temperature analysis from EIS in Section \ref{sec:EIStemp} is also in line with small temperature changes across the coronal hole. Therefore, we do not expect a strong, large-scale spatial-temporal temperature gradient in the coronal hole examined in this study, and attribute Mg/Ne FIP bias variability in Figure \ref{fig:solo_remote_data_SPICE} in large part to changes in its compositional properties. Some variation in high-FIP Ne has been observed solar minimum expressed remotely \citep{Landi2015} and in situ \citep{Shearer2014}, however we do not expect these effects to drive changes in the SPICE observations at solar maximum.

Through this implementation, we find the FIP bias of low-FIP Mg and high-FIP Ne across the raster falls within 1--6. A large portion of the coronal hole is between 1--2 while the edges extending to higher FIP values. The {\alf}ic slow wind footpoints correspond to inside of the coronal hole boundary within 1--2 FIP bias values that would coincide with a low-FIP to high-FIP in situ ratio, Fe/O FIP bias, $\sim 1$ at Solar Orbiter, shown in panel L of Figure \ref{fig:Parker_solo_insitu_data}. However, we note that the FIP bias map at the Sun shows a range of photospheric ($\sim1$) and low-FIP enhanced regions within and outside of the coronal hole making it difficult to confirm footpoints with certainty. We also note that the FIP bias comparison between the remote observations and in situ includes ratios from different elements that may undergo different degrees of the FIP effect, therefore the connection between Mg/Ne (remote) and Fe/O (in situ) may not be directly comparable \citep{Laming2019}. Overall, a more rigorous comparison of elemental abundances requires several and matching elemental ratios between remote and in situ heavy ions \citep{Rivera2022}. 

\subsection{Electron Temperature and Density} \label{sec:EIStemp}
Given that EIS observes lines from more highly ionized ions in the corona, we use those observations to determine a coronal temperature and density associated near the edge of the coronal hole boundary to compare with the isothermal radius modeling discussed in Section \ref{sec:polytropic}. Figure \ref{fig:EIS_rmote_data} shows the EIS intensity (left) and Doppler shifts (right) map of the Fe {\footnotesize XII} 195.119\AA\/ spectral line with the coronal hole outline overlayed. Using the EIS Python Analysis Code (EISPAC) software package \citep{Weberg_eispac}, the intensity is computed by integrating over the fitted Gaussian profile of the Fe {\footnotesize XII} spectral line. The Doppler shifts are computed as the shift of the fitted line centroid from the theoretical emission line peak in km s$^{-1}$. A red doppler shift shows plasma moving away from the observer while a blue shift shows plasma moving towards the observer. 

We identify four small $20\times20$ arcsecond regions near the edge and inside of the coronal hole, shown as purple, blue, green, and red boxes in Figure \ref{fig:EIS_rmote_data}, that would correspond to the {\alf} slow solar wind footpoints. The plasma is assumed to be uniform in temperature and density within each of the boxes. To determine the electron density, we use the standard density sensitive line pair, Fe {\footnotesize XIII} 202.044/203.826, where the intensity ratio simplifies to the ratio of their contributions functions \citep{Brooks2011, Brooks2022}. We list the lines and their intensities in Table \ref{table:EIS_intensities}. The average electron density within purple, blue, green, and magenta boxes are $5.9\times10^8$, $1.6\times10^8$, $4.6\times10^8$, $9.8\times10^8$, cm$^{-3}$ respectively. We include these estimates in the solar wind modeling results as a constraint to the proton density in comparison to the initial density, under the assumption of quasi-neutrality holds.

Traditionally, EIS observations provide lines from several consecutive Fe ions to derive a DEM of the observed plasma as a function of temperature. A well constrained DEM curve can be used to determine an electron temperature. However, the lines observed during the raster in the present study are limited due to a noisy signal inside coronal hole boundary. Instead, we estimate the temperature using a line ratio of consecutive ions using a sum of the Fe {\footnotesize XI} 188.216 + Fe {\footnotesize XI} 188.299 + Fe {\footnotesize XI} 192.813 to the sum of Fe {\footnotesize XII} 195.119 + Fe {\footnotesize XII} 192.394 to estimate an electron temperature for each box, as listed in Table \ref{table:EIS_intensities}. Figure \ref{fig:EIS_temps} shows the range of temperatures for each corresponding box, indicating a source temperature of $\sim 1.17-1.19$ MK. We include this estimate solar wind modeling results as a constraint to the electron temperature.

\section{Radial Profiles and Parker Solar Wind Model Comparison}\label{sec:polytropic}

Following \cite{Rivera2024} we implement two fluid,``iso-poly'' Parker solar wind models \citep{Dakeyo2022} and implement an external fitted wave force based on two point measurements of the {\alf} wave pressure. These measurements are used to constrain an analytic formula \citep{Shi2022} which is essentially a power law between the two measured points and parameterized by an amplitude $f_0$ in units of $GM_\odot/R_\odot^2$ and a power law index $\nu$ \citep[see][ for more details]{Rivera2024}.  

\begin{figure*}[]
	\centering
	\includegraphics[width=\linewidth]{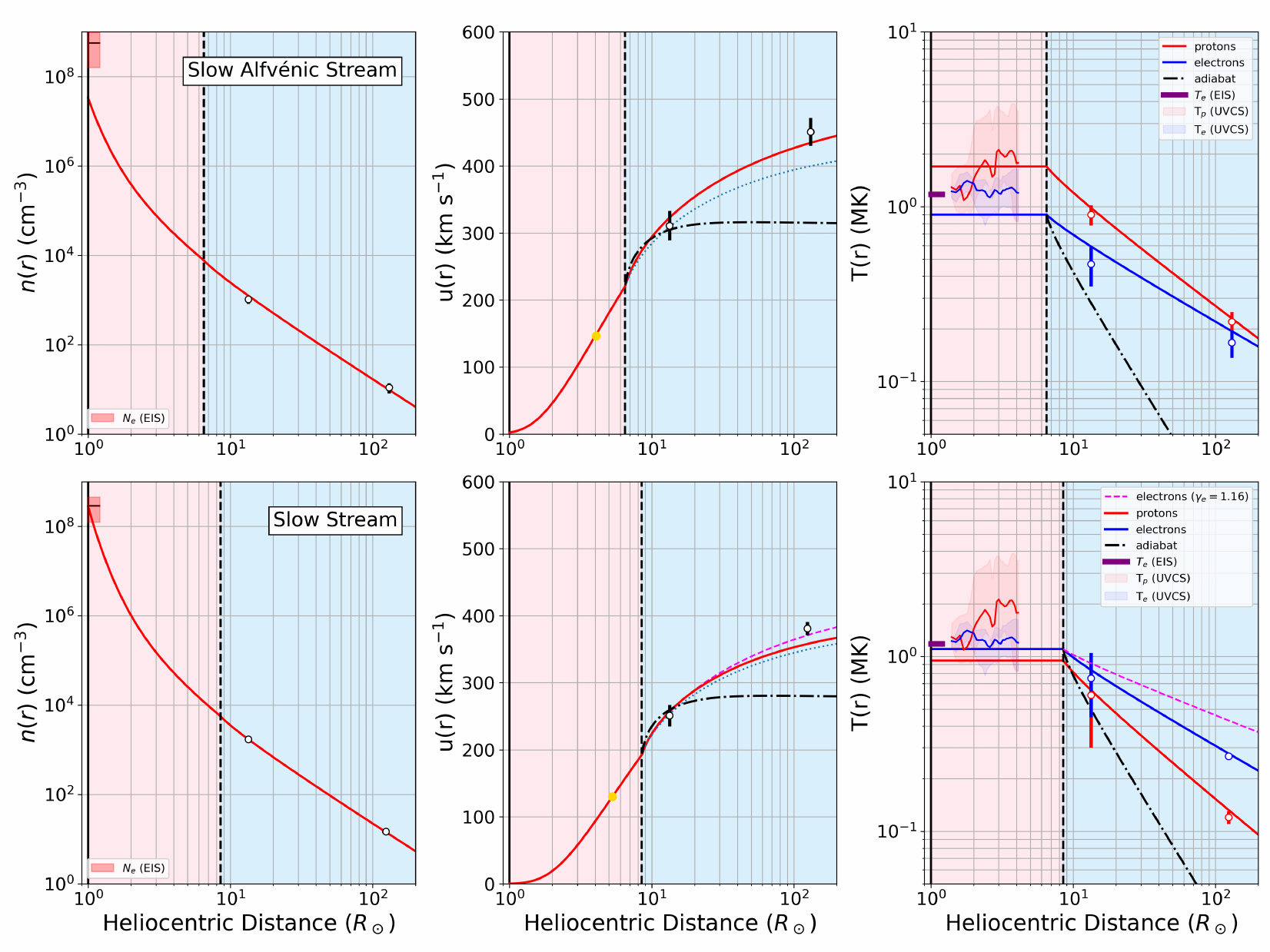}
	\caption{Parker solar wind solutions and associated data constraints for the Alfv\'enic slow (top row) and slow (bottom row). From left to right: Proton density, proton velocity, proton (red) and electron (blue)  temperatures. For the acceleration profiles (middle column), solutions are shown both with (solid) and without (dotted) Alfv\'enic wave pressure as well as for the adiabatic/non-Alfv\'enic wave pressure case (dashed-dotted). For the slow stream (bottom row) an additional solution is shown with the electron polytropic index decreased (made closer to isothermal) to illustrate the magnitude of the change needed to explain the observed acceleration (dashed magenta). The white circles are the average properties at Parker and Solar Orbiter for the slow wind in Table \ref{table:CharacteristicPropertiesSlow} and {\alf}ic slow wind in Table \ref{table:CharacteristicPropertiesAlfvenicSlow}. We also include remote observations of coronal density and electron temperature derived in Section \ref{sec:EIStemp} from EIS and the electron and proton temperature from previously examined coronal holes from UVCS observations over the poles during the 1996–1997 \cite{Cranmer2020a}.}
	\label{fig:parker_solutions}
\end{figure*}

This modeling approach solves the 1D radial expansion of a hydrodynamic fluid under the joint effects of outwards thermal pressure gradients from electrons and protons, the aforementioned wave pressure gradient and the inward force of gravitation. The results are shown in figure \ref{fig:parker_solutions} where the models and data constraints on density, velocity and temperature (for both protons and electrons) are plotted.

The models couple two distinct thermal regimes~: an isothermal corona \cite{Parker1958} and a polytropic ($\gamma < 5/3$) inner heliosphere \cite{Parker1960} in the young solar wind. Varied parameters are the average isotropic temperatures of the protons and electrons in the corona, the polytropic index of both protons ($\gamma_p$) and electrons ($\gamma_e$), and the radius at which the thermal regime transition from corona to inner heliosphere ($R_{iso}$). These parameters set the temperature profiles (right hand column, figure \ref{fig:parker_solutions}) and the acceleration profiles are then prescribed (middle hand column, Figure \ref{fig:parker_solutions}). The density curve follows mass flux conservation (left hand column, figure \ref{fig:parker_solutions}) and is matched to measurements by varying a reference density ($n_{0p}$) at $1 R_\odot$. 

Measured proton temperatures at both spacecraft provide a strong constraint on the proton polytropic index, $\gamma_p$. For the electron temperatures we obtain measurements at Parker from \citet{Halekas2020}, while in lieu of a Solar Orbiter measurement, we use the statistically derived electron polytropic index most appropriate for the observed wind speeds from \citet{Dakeyo2022}. The temperature constraints also couple $R_{iso}$ and the coronal temperatures (T$_{p0}$,T$_{e0}$): since the derived temperature profiles are fitted to the observations in the interplanetary medium, decreasing $R_{iso}$ increases the necessary coronal temperature. These three parameters are therefore varied together to find a solution which best matches the observed acceleration of the stream while remaining consistent with coronal constraints. 

The end results are the modeling parameters recorded in table \ref{table:ModelParams}. In figure \ref{fig:parker_solutions}, we plot the resulting solutions both with (dotted) and without (solid) the effect of the observed wave energy flux, as well as the solutions for the adiabatic case in dashed dotted lines. We also include coronal proton and electron measurements determined remotely from UltraViolet Coronagraph Spectrometer (UVCS) computed in \cite{Cranmer2020a} and from the analysis of EIS observations in Section \ref{sec:EIStemp} to compare with the resulting isothermal layer extrapolations. The UVCS temperatures were computed for polar coronal holes between 1996–1997 and most representative of fast wind conditions. For comparison to the modeled density profile, we include the computed electron density (from EIS) as the proton density at the solar surface.

\begin{table*}
\centering
\begin{tabular}{cccccccc}
	\hline
	
Stream & n$_{p0}$ (cm$^{-3}$) & T$_{p0}$ (MK) & T$_{e0}$ (MK) & $\gamma_{p}$ & $\gamma_{e}$ & R$_{iso}$ ($R_{\odot}$) & [f$_0$ (GM$_{\odot}$/$R_{\odot}^2$), $\nu$] \\ \hline
Slow Alfv\'enic &3.3$\times10^7$&1.7&0.9&1.3&1.23&6.5&[3.7$\times 10^{-4}$,1.5] \\ \hline
Slow            &$2.3\times10^8$&0.95&1.1&1.33&1.23&8.5&[4.1$\times 10^{-4}$,1.1] \\ \hline
\end{tabular} \\
\caption{Parameters for the solar wind models shown in Figure \ref{fig:parker_solutions}.}
\label{table:ModelParams}
\end{table*}

For the {\alf}ic slow wind stream (top row), the observed thermal pressure gradients (within the temperature measurement errorbars) and wave energy flux can explain the observed acceleration. Moreover, the wave energy flux is seen to be important for this stream, similar to that reported in \cite{Rivera2024} for fast wind. Without it, the accelerated stream only reaches 400km s$^{-1}$ at Solar Orbiter but with it the observed speed reaches within error bars of the $451\pm21$km s$^{-1}$ measurement. In the adiabat case, the acceleration  beyond Parker is negligible. We note that even with both the thermal and wave pressures (and the thermal pressure near the top of the measurement error bars) the achieved acceleration is still slightly below the mean measurement. This may indicate other contributions, for example the effect of stream interactions, are needed to fully account for the stream evolution.

Meanwhile for the slow stream (bottom row), the effect of wave energy flux is negligible \citep[see ][]{Halekas2022,Halekas2023}. However, no combination of parameters which matched the temperature observations was found to fully explain the observed acceleration. Decreasing the electron polytropic index by 6\% of the statistical value from \citet{Dakeyo2022} for a $\gamma_e =1.16$ quantifies the additional pressure needed to solve the gap (magenta dashed curves), but other processes such as transverse density gradients or non-{\alf}ic wave pressure may play a role given that {\alf}ic wave pressure cannot explain this gap. 

We also note that while proton thermal pressure dominates over electron thermal pressure in the slow {\alf}ic case, the inverse is true for the slow wind case in our study. In all regards, the {\alf}ic slow wind stream here qualitatively resembles the properties of fast wind, consistent with many prior studies \citep{DAmicis2015, DAmicis2019, DAmicis2021, Ervin2024b}.

\section{Discussion} \label{sec:discussion}
The work maps a single stream of slow and {\alf}ic solar wind from the Sun to the inner heliosphere to trace large-scale energetics within their radial evolution. Both streams are intercepted by Parker ($\sim13R_{\sun}$) and Solar Orbiter ($\sim130R_{\sun}$) where the streams are matched by comparing magnetic field polarity, compositional properties, and conserved quantities based on their ballistically backmapped source surface footpoints. A combination of mass and magnetic flux, and the total energy flux (Kinetic, Enthalpy, Gravitational, Wave) is conserved between their heliocentric locations within the variability of the stream properties. Energy flux conservation suggests that the energy within each stream is exchanged self-consistently across the components.

Through a comparison of the energy flux budgets of the slow and {\alf}ic slow wind, we find that different energy components drive their respective heating and acceleration through the inner heliosphere. The {\alf}ic slow solar wind contains a larger relative wave energy flux contribution to the total energy flux ($\sim7\%$) compared to the non-{\alf}ic slow ($\sim2\%$) at Parker. The wave energy flux is found to decay to $\sim2.3\%$ ({\alf}ic) and $\sim0.4\%$ ({non-\alf}ic) of the total energy flux by Solar Orbiter. In comparison, a fast wind stream from a similar period and heliocentric evolution (from the blue shaded region in Figure \ref{fig:Parker_solo_insitu_data}) is found to contain a larger relative wave energy flux contribution of $\sim10\%$ that decreases to $\sim1\%$ of the total energy budget between Parker and Solar Orbiter \citep{Rivera2024}. In the case of the enthalpy contribution at Parker, the slow wind stream is dominated by the electron population, where T$_e$/T$_p = 1.25$, while this ratio is only about 0.5 in the {\alf}ic slow wind. In comparison to the fast wind case, the temperature ratio at Parker is found to be smaller than both slow wind cases, T$_e$/T$_p =0.35$ \citep{Rivera2024}. In both the slow wind cases, the decrease in the enthalpy and wave energy flux is balanced by an increase in the kinetic and gravitational terms, such that energy is conserved, as visualized in Figure \ref{fig:energy_fluxes}. We note that the electron heat flux has a larger relative contribution at Parker in the slow versus {\alf}ic slow wind stream and therefore it is included in the energy budget. However, the unavailable electron measurements at Solar Orbiter inhibited a direct computation of the electron heat flux there. Instead, we use the same $\sim4\%$ contribution as was found at Parker distances, in accordance to statistical results from \cite{Halekas2023}. Therefore, the electron heat flux in the slow wind stream is expected to remain at $4\%$.

Overall, the two fluid, iso-poly + wave forcing Parker solar wind modeling results show that the observed wave energy flux and associated wave pressure gradient is a necessary component to explain the full {\alf}ic slow wind acceleration, in addition to the non-adiabatic proton thermal pressure gradient. In contrast, the slow wind's acceleration is largely accounted for by the non-adiabatic thermal pressure gradient of the electrons, with a smaller proton contribution, without a significant contribution from the wave pressure gradient. We note that while the waves are not contributing to the direct acceleration of the solar wind in this stream, their dissipation may still be relevant to producing the non-adiabatic pressure gradient, which can be driven by ion-cyclotron waves and turbulent dissipation in streams around the HCS \citep{Telloni2023b}. A summary of the results for the different cases of the velocity profile is shown in the middle panel of Figure \ref{fig:parker_solutions}. Both the slow and {\alf}ic slow wind proton temperature profiles have a similar non-adiabatic cooling profile, with a fitted polytropic index of 1.3 ({\alf}ic) and 1.33 (non-{\alf}ic). We find the two slow wind cases exhibit a shallower proton temperature profile compared to the fast wind case of \cite{Rivera2024}, showing $\gamma_p=1.4$. Because of the missing electron temperature at Solar Orbiter, we instead employ an electron polytropic index, $\gamma_e$, from the associated wind speed family of \cite{Dakeyo2022}. However, we note that the electron temperature profile (blue curve in Figure \ref{fig:parker_solutions}) is not able to fully meet the acceleration exhibited by the stream. We explore the excess forcing needed to resolve this discrepancy. We find that by making the electron temperature curve $6\%$ shallower (smaller polytropic index) than statistical trends, the observed acceleration can be achieved (magenta dashed curve). This entails adding 2 W m$^{-2}$ (roughly $4\%$ of the total energy flux) to the energy budget. It is interesting to note this is comparable to the $1.75$ W m$^{-2}$ present in the electron heat flux at Parker, therefore if a substantial portion of this was transferred to the protons during transport it could contribute to this necessary extra forcing while keeping the electron polytropic index more in line with expectations.

Finally, we find the extrapolated temperature and density simulated curves in the corona are somewhat compatible with remote sensing constraints of the {\alf}ic solar wind source region while highly compatible with the slow wind solutions. The solar wind modeling is further constrained through remote observations from Hinode/EIS where the electron temperature and density were derived for the coronal hole structure associated with the {\alf}ic slow solar wind. We also include proton and electron temperature (spanning $1.5-4R_{\sun}$) from a polar coronal hole from \cite{Cranmer2020a} to compare with our temperature results. Given that the Parker iso-poly solar wind solutions are derived completely independently from any coronal observations, their relative agreement lends validity to the resulting stream profile in the isothermal layer. However, the discrepancy observed is likely due to the complexity of the corona itself and the fact that EIS only provides a single height of coronal observations along the LOS. Alternatively, the simulated isothermal temperature region uses the same radius for both proton and electron temperature profiles which are likely independent. Adjusting the isothermal radius separately could lead to different isothermal temperature results that may be more consistent with the heliospheric electron profile \citep{Dakeyo2022}. Also, we find that the UVCS coronal observations from polar coronal holes are most compatible with the {\alf}ic slow solar wind coronal observations from EIS in line with a coronal hole source region, as shown in Figure \ref{fig:solo_remote_data_fulldisk}. 

Ideally, an off-limb, radial profile (as in the UVCS case) of the proton and electron temperature and density, along with the bulk solar wind speed, including plasma composition, with magnetic field and {\alf} wave properties along the individual streams would provide more rigorous constraints to plasma source region footpoints and evolution below Parker's orbit. This would only possible in cases where remote observations (including polarimetric observations that inform on the magnetic field morphology, e.g. \citealt{Yang2024, Schad2024}) of the corona are taken in quadrature with in situ observations of the connected source region footpoints. Future work will examine such a spacecraft alignment to capture more of the extended coronal properties to constrain solar wind energetics across the middle corona \citep{delZanna2018, West2023}. 

\section{Summary and Conclusions}  \label{sec:conclusions}
Through linked remote and in situ observations of a slow and {\alfic} slow solar wind stream, we follow their propagation from the corona through the inner heliosphere. A two-fluid (protons and electrons) implementation of the Parker solar wind solutions with an added wave forcing term, constrained to observations at two heliocentric distances and previous statistical studies, generate the appropriate thermal and wave pressure gradients that explain the observed heating and acceleration of the {\alf}ic slow solar wind stream. In contrast, the solutions for the slow solar wind acceleration is generated by the electron and proton thermal pressure gradients, without a significant contribution from {\alf} waves. However, we note that the slow wind requires a slightly shallower electron polytropic index compared to a statistically derived value from a similar asymptotic wind speed profile (using 94\% of the stated value from \citealt{Dakeyo2022} from solar wind family C.) to reach the observed speed at Solar Orbiter.

A comparison between the slow, {\alfic} slow, and a previously studied fast wind stream \citep{Rivera2024}, from the same period, indicates an increasing contribution of the wave energy flux to the energy budget, that is in line with statistical work from \citealt{Halekas2023}. The increasing wave energy flux is directly correlated to wind speed, indicating a growing importance to the acceleration of the solar wind. The details of how the waves are converted to heat warrant further investigation; they are likely to involve reflection driven turbulence, e.g. \citealt{Chandran2009, vanBallegooijen2016}, and turbulent dissipation, e.g. \citealt{Sorriso-Valvo2007, MacBride2008, Stawarz2009}. Previous work has shown that turbulent cascade rates (i.e., the rate at which turbulent dynamics bring energy to the small scales to facilitate dissipation) are consistent with the heating rates needed to produce the radial temperature profiles observed in the solar wind. This likely has a direct relationship to the decay of switchbacks discussed in this study \citep{Marino2008, Stawarz2009, Coburn2012, Bandyopadhyay2023, Bourouaine2024}.

Coronal constraints of temperature and density are somewhat compatible with those derived independently with the modeling but more extended observations of the corona are required fully constrain the solar wind evolution below Parker's orbit. Previous work has shown great progress in accessing the solar and solar wind connection using remote observations, magnetohydrodynamic (MHD) simulations, and in situ measurements from orbiting spacecraft \citep{Parenti2021, Knizhnik2024} with many implementing several points of constraint across the Sun-Heliospheric system to connect coronal outflows to the inner heliosphere \citep{Brooks2021, Baker2023, Telloni2021, Adhikari2022}. Future work will leverage coordinated remote and in situ observations to tracing energetics from deep in the corona to the solar wind.

Y.J.R. is partially supported by the Parker Solar Probe project through the SAO/SWEAP subcontract 975569. 
J.E.S. is supported by the Royal Society University Research Fellowship URF/R1/201286. 
B.L.A. acknowledges NASA contract and NASA Grants 80NSSC22K0645 (LWS/TM), 80NSSC22K1011 (LWS), and 80NSSC20K1844.
K.K.R is funded by the NASA HSO Connect program, grant number 80NSSC20K1283. Special thanks to Peter Young and Teodora Mihailescu for thoughtful discussion on the Hinode/EIS observations.

Parker Solar Probe was designed, built, and is now operated by the Johns Hopkins Applied Physics Laboratory as part of NASA’s Living with a Star (LWS) program (contract NNN06AA01C). Support from the LWS management and technical team has played a critical role in the success of the Parker Solar Probe mission. Solar Orbiter is a mission of international cooperation between ESA and NASA, operated by ESA. Solar Orbiter SWA data were derived from scientific sensors that were designed and created and are operated under funding provided by numerous contracts from UKSA, STFC, the Italian Space
Agency, CNES, the French National Centre for Scientiﬁc
Research, the Czech contribution to the ESA PRODEX
program, and NASA.  The SWA-HIS team acknowledges NASA contract NNG10EK25C.  Solar Orbiter SWA work at the UCL/Mullard
Space Science Laboratory is currently funded by STFC (grant
Nos. ST/W001004/1 and ST/X/002152/1).
The SWA team at INAF/IAPS is currently funded under ASI grant 2018-30-HH.1-2022.
{\em Hinode} is a Japanese mission developed and launched by ISAS/JAXA, with NAOJ as domestic partner and NASA and STFC (UK) as international partners. It is operated by these agencies in co-operation with ESA and NSC (Norway).

\bibliography{refs.bib}

\end{document}